\documentclass[12pt]{article}
\def\ket#1{\left| #1 \right\rangle}
\def\vev#1{\left\langle #1 \right\rangle}
\def\no#1{{\rm :} \,  #1 \, {\rm :}}
\begin{document}
\baselineskip=15.5pt
\renewcommand{\theequation}{\arabic{section}.\arabic{equation}}
\pagestyle{plain}
\setcounter{page}{1}
\def\appendix{
\par
\setcounter{section}{0}
\setcounter{subsection}{0}
\def\thesection{\Alph{section}}}
\begin{titlepage}

\rightline{\tt hep-th/0311115}
\rightline{\small{\tt MIT-CTP-3440}}
\rightline{\small{\tt CALT-68-2462}}

\begin{center}

\vskip 1.8cm

{\LARGE {Solving Witten's string field theory}}

\vskip 0.5cm

{\LARGE{using the butterfly state}}  

\vskip 1.8cm

{\large {Yuji Okawa}}

\vskip 0.5cm

{\it {Center for Theoretical Physics, Room 6-304}}

\smallskip

{\it {Massachusetts Institute of Technology}}

\smallskip

{\it {Cambridge, MA 02139, USA}}

\smallskip

okawa@lns.mit.edu

\vskip 1.8cm

{\bf Abstract}
\end{center}

\noindent
We solve the equation of motion
of Witten's cubic open string field theory
in a series expansion using the regulated butterfly state.
The expansion parameter is given by the regularization parameter
of the butterfly state, which can be taken to be arbitrarily small.
Unlike the case of level truncation,
the equation of motion can be solved
for an arbitrary component of the Fock space
up to a positive power of the expansion parameter.
The energy density of the solution is well defined
and remains finite even in the singular butterfly limit,
and it gives approximately 68\% of the D25-brane tension
for the solution at the leading order.
Moreover, it simultaneously solves the equation of motion
of vacuum string field theory,
providing support for the conjecture at this order.
We further improve our ansatz by taking into account
next-to-leading terms, and find two numerical solutions
which give approximately 88\% and 109\%, respectively, of
the D25-brane tension for the energy density.
These values are interestingly close to those
by level truncation at level 2
without gauge fixing studied
by Rastelli and Zwiebach and by Ellwood and Taylor.

\end{titlepage}

\newpage

\section{Introduction and summary}
\setcounter{equation}{0}

Ever since the pioneering work of Sen and Zwiebach \cite{Sen:1999nx},
it has been demonstrated
that Witten's cubic open string field theory \cite{Witten:1985cc}
is capable of describing nonperturbative phenomena
such as tachyon condensation
\cite{Moeller:2000xv, deMelloKoch:2000ie, Moeller:2000jy,
Taylor:2002fy, Gaiotto:2002wy}.\footnote
{
For a recent review, see \cite{Taylor:2003gn}.
}
The action of Witten's string field theory
is given by \cite{Witten:1985cc}
\begin{equation}
  S = -\frac{1}{\alpha'^3 g_T^2} \left[
  \frac{1}{2} \vev{ \Psi | Q_B | \Psi }
  + \frac{1}{3} \vev{ \Psi | \Psi \ast \Psi } \right],
\label{Witten-action}
\end{equation}
where $Q_B$ is the BRST operator and $g_T$
is the on-shell three-tachyon coupling constant,
and its equation of motion is
\begin{equation}
  Q_B \ket{\Psi} + \ket{ \Psi \ast \Psi } = 0.
\end{equation}

The tachyon potential of Witten's string field theory
calculated by an approximation scheme called level truncation
\cite{Kostelecky:1989nt, Sen:1999nx, Moeller:2000xv,
Taylor:2002fy, Gaiotto:2002wy}
has reproduced the D25-brane tension
with impressive precision,
providing strong evidence
for Sen's conjecture \cite{Sen:1999xm, Sen:1999mh}.
For instance, if we truncate the string field $\ket{\Psi}$
up to level 2 in the Siegel gauge,
\begin{equation}
  \ket{\Psi}
  = x \, c_1 \ket{0} + 2 \, u \, c_{-1} \ket{0}
    + v \, L^m_{-2} \, c_1 \ket{0},
\end{equation}
the absolute value of the energy density of the solution
obtained by taking variations
with respect to the parameters $x$, $u$, and $v$
is given by approximately 95.9\%
of the D25-brane tension \cite{Sen:1999nx}.
However, there are infinitely many other components
in the string field so that the solution by level truncation
does not solve the equation of motion for a generic component
above the truncation level.

On the other hand, various exact solutions of string field theory
based on the identity state have been constructed and studied
\cite{Horowitz:dt, Takahashi:2001pp, Kishimoto:2001de,
Kluson:2002ex, Takahashi:2002ez, Kluson:2002hr, Kishimoto:2002xi,
Kluson:2002gu, Kluson:2002te, Kluson:2002av, Takahashi:2003pp,
Kluson:2003xu, Takahashi:2003xe}.
They exactly solve the equation of motion when it is contracted
with any state in the Fock space,
but when we try to evaluate the energy density for the solutions,
we encounter the notorious singularity coming from
the inner product of the identity state with itself.

These previous attempts to solve the equation of motion
thus have both good and bad features, which can be
summarized in the following way.
We expect that the analytic solution $\ket{\Psi}$ corresponding to
the tachyon vacuum satisfies the following conditions:
\begin{quotation}
\noindent
(i)  It solves the equation of motion,
\begin{equation}
  \vev{ \phi | Q_B | \Psi } + \vev{ \phi | \Psi \ast \Psi } = 0,
\end{equation}
for any state $\ket{\phi}$ in the Fock space.\\

\noindent
(ii) The quantities $\vev{ \Psi | Q_B | \Psi }$ and
$\vev{ \Psi | \Psi \ast \Psi }$ are well defined, and\\
\noindent
(ii-a) the equation of motion is satisfied even when
contracted with the solution itself:
\begin{equation}
  \vev{ \Psi | Q_B | \Psi } + \vev{ \Psi | \Psi \ast \Psi } = 0,
\end{equation}
(ii-b) the energy density ${\cal E}$ of the solution
is given by the D25-brane tension $T_{25}$ as follows
\cite{Sen:1999xm, Ghoshal:2000gt}:
\begin{equation}
  \int d^{26} x \, {\cal E} [ \Psi ]
  = \frac{1}{\alpha'^3 g_T^2} \left[
  \frac{1}{2} \vev{ \Psi | Q_B | \Psi }
  + \frac{1}{3} \vev{ \Psi | \Psi \ast \Psi } \right], \quad
  \frac{{\cal E} [ \Psi ]}{T_{25}} = -1,
\end{equation}
where $T_{25}$ is expressed in terms of $g_T$ by\footnote
{
See Appendix A of \cite{Okawa:2002pd} for a check of this relation
following Polchinski's conventions in \cite{Polchinski:rq}.
}
\begin{equation}
  T_{25}= \frac{1}{2 \pi^2 \alpha'^3 g_T^2}.
\end{equation}
\end{quotation}
The solution by level truncation satisfies (ii-a)
by construction, and satisfies (ii-b) with very high precision,
while the condition (i) is not satisfied
for an arbitrary $\ket{\phi}$.
The exact solutions based on the identity state
satisfy the condition (i), but the quantities
$\vev{ \Psi | Q_B | \Psi }$ and $\vev{ \Psi | \Psi \ast \Psi }$
are not well defined so that we cannot even address the questions
whether the conditions (ii-a) and (ii-b) are satisfied.

In this paper, we present a new approach to this problem.
Our solution satisfies all the above conditions
approximately but simultaneously.
Our ansatz for a solution
takes the form of the regulated butterfly state
\cite{Gaiotto:2001ji, Schnabl:2002ff, Gaiotto:2002kf, Okawa:2003cm}
with an operator inserted at the midpoint of the boundary.\footnote
{
An analytic framework to solve
the equation $(L_0-1) \ket{\Psi} + \ket{ \Psi \star \Psi } = 0$
using the butterfly state was proposed in \cite{Bars:2003cr}.
Here $\star$ is the Moyal star product
\cite{Bars:2001ag, Bars:2002bt, Bars:2002nu,
Bars:2002qt, Bars:2002yj, Bars:2003cr, Bars:2003gu},
and a closed expression of a formal exact solution has been
provided as a perturbative expansion.
We use the butterfly state in a different way.
Our expansion scheme and lowest-order solution
are different from those of \cite{Bars:2003cr}.
}

The butterfly state \cite{Gaiotto:2001ji}
is a star-algebra projector,
and the regulated butterfly state
\cite{Schnabl:2002ff, Gaiotto:2002kf} is its regularization.
It is parametrized by $t$ in the range $0 \le t < 1$,
and the butterfly state is given by the limit $t \to 1$.
Our solution satisfies the condition (i)
up to a positive power of $1-t$.
The solution at the leading order is given by
choosing a $c$ ghost as the operator to be inserted.
If we denote the regulated butterfly state
with the operator ${\cal O}$ inserted by $\ket{B_t(\cal O)}$,
the solution takes the following form:
\begin{equation}
  | \Psi^{(0)} \rangle = \frac{x}{\sqrt{1-t}} \ket{B_t(c)},
\end{equation}
where $x$ is a parameter to be determined.
We will show that the equation of motion can be solved
up to $O(\sqrt{1-t})$:
\begin{equation}
  \langle \phi | Q_B | \Psi^{(0)} \rangle
  + \langle \phi | \Psi^{(0)} \ast \Psi^{(0)} \rangle
  = O(\sqrt{1-t})
\end{equation}
for any state $\ket{\phi}$ in the Fock space by this ansatz.
The operator $c$ inserted into the regulated butterfly state
corresponds to the state $c_1 \ket{0}$
at level 0. If we incorporate operators
corresponding to states at level 2 into our ansatz
\begin{eqnarray}
  | \Psi^{(2)} \rangle &=& \frac{x}{\sqrt{1-t}} \ket{B_t(c)}
\nonumber \\
  && {} + \sqrt{1-t} \left[ u \, | B_t(\partial^2 c) \rangle
  + v \, \ket{ B_t(c T^m) }
  + w \, \ket{ B_t(\no{ bc \partial c}) } \right],
\label{Psi^(2)}
\end{eqnarray}
where $x$, $u$, $v$, and $w$ are parameters to be determined,
it turns out that the equation of motion can be
solved up to a higher power of $1-t$:
\begin{equation}
  \langle \phi | Q_B | \Psi^{(2)} \rangle
  + \langle \phi | \Psi^{(2)} \ast \Psi^{(2)} \rangle
  = O((1-t)^{\frac{3}{2}})
\end{equation}
for any state $\ket{\phi}$ in the Fock space.
We find one analytic solution in the form of $| \Psi^{(0)} \rangle$
and two numerical solutions $| \Psi^{(2)}_1 \rangle$ and
$| \Psi^{(2)}_2 \rangle$ in the form of $| \Psi^{(2)} \rangle$.
Therefore, the condition (i) can be satisfied
with high precision if we choose $t$ to be close to $1$.
Furthermore, it turns out that
the inner products appearing in the condition (ii)
are well defined for our solutions.
In order to discuss the condition (ii-a),
let us define the ratio ${\cal R} [ \Psi ]$
for a string field $\ket{\Psi}$ by
\begin{equation}
  {\cal R} [ \Psi ] \equiv
  \frac{\langle \Psi | Q_B | \Psi \rangle}
  {\langle \Psi | \Psi \ast \Psi \rangle}.
\label{R-definition}
\end{equation}
The condition (ii-a) is satisfied when ${\cal R} [ \Psi ] = -1$.
For the leading-order solution $| \Psi^{(0)} \rangle$,
${\cal R} [ \Psi^{(0)} ]$ and ${\cal E} [ \Psi^{(0)} ] / T_{25}$
are given by
\begin{eqnarray}
  && \lim_{t \to 1} {\cal R} [ \Psi^{(0)} ]
  = -2^{\frac{15}{2}} \cdot 3^{-\frac{19}{4}} \simeq -0.9804,
\nonumber \\
  && \lim_{t \to 1} \frac{{\cal E} [ \Psi^{(0)} ]}{T_{25}}
  = 2 \pi^2 (-2^{-4} \cdot 3^{\frac{1}{2}}
             +2^{-\frac{21}{2}} \cdot 3^{\frac{17}{4}})
    \simeq -0.6838.
\end{eqnarray}
The condition (ii-a) is satisfied
with precision of approximately 98\%,
and the absolute value of the energy density
gives approximately 68\% of the D25-brane tension.
These values are improved for
the solutions at the next-to-leading order:
\begin{eqnarray}
  && \lim_{t \to 1}
  {\cal R} [ \Psi^{(2)}_1 ] \simeq -0.9989, \quad
  \lim_{t \to 1} \frac{{\cal E} [ \Psi^{(2)}_1 ]}{T_{25}}
  \simeq -0.8826,
\\
  && \lim_{t \to 1}
  {\cal R} [ \Psi^{(2)}_2 ] \simeq -0.9981, \quad
  \lim_{t \to 1} \frac{{\cal E} [ \Psi^{(2)}_2 ]}{T_{25}}
  \simeq -1.0898.
\end{eqnarray}
The condition (ii-a) is satisfied with better precision than 99.8\%,
and the values of the energy density for these solutions are
approximately 88\% and 109\%, respectively, of the D25-brane tension.

It is not surprising that there are more than one solution
at this order since we do not fix gauge.
There are also two solutions at level $2$ in level truncation
without gauge fixing studied
by Rastelli and Zwiebach \cite{Rastelli:2000iu}
and by Ellwood and Taylor \cite{Ellwood:2001ne}.
In fact, the values of the ratio ${\cal E}/T_{25}$
for our solutions are interestingly close to those from
the level-truncation analysis without gauge fixing, which are
${\cal E}/T_{25} = -4096 \pi^2/59049 \simeq -0.6846$
for the solution at level $0$, and
${\cal E}/T_{25} \simeq -0.8802, \, -1.0778$
for the two solutions at level $2$ \cite{Ellwood:2001ne},
as compared in Table \ref{comparison}.

\begin{table}[ht]
\caption{Comparison of $-{\cal E}/T_{25}$
with level truncation without gauge fixing.}
\label{comparison}
\begin{center}
{\renewcommand\arraystretch{1.3}
\begin{tabular}{|c|c||c|c|}
  \hline
  Our solutions & $-{\cal E}/T_{25}$
  & Level truncation & $-{\cal E}/T_{25}$ \\
  \hline
  $\Psi^{(0)}$ & $0.6838$ & level 0 & $0.6846$ \\
  \hline
  $\Psi^{(2)}$ & $0.8826$, $1.0898$ & level 2 & $0.8802$, $1.0778$ \\
  \hline
\end{tabular}
}
\end{center}
\end{table}

There is another interesting aspect of our solution.
Our leading-order solution $| \Psi^{(0)} \rangle$ satisfies
not only the equation of motion of Witten's string field theory
but also that of vacuum string field theory
\cite{Rastelli:2000hv, Rastelli:2001jb, Rastelli:2001uv}
at the leading order,
\begin{equation}
  - \langle \phi | {\cal Q} | \Psi^{(0)} \rangle
  + \langle \phi | \Psi^{(0)} \ast \Psi^{(0)} \rangle
  = O(\sqrt{1-t}),
\end{equation}
when ${\cal Q}$ is given by the midpoint $c$-ghost insertion
with an appropriate normalization \cite{Gaiotto:2001ji}.
This can be regarded as support
for the vacuum string field theory conjecture at this order.
It was shown \cite{Gaiotto:2001ji} that
a star-algebra projector in the twisted ghost conformal
field theory solves the equation of motion of
vacuum string field theory in the singular limit,
but our solution is different from the twisted butterfly state
and it is a new solution to vacuum string field theory.
Note also that the twisted butterfly state does not solve
the equation of motion of Witten's string field theory
even at the leading order.

It is not difficult to see how our ansatz works.
Actually, we can use any regulated star-algebra projector
instead of the regulated butterfly state.\footnote
{
We are aware that this idea at the leading order
occurred to many others independently including I.~Ellwood,
L.~Rastelli, M.~Schnabl, A.~Sen, and B.~Zwiebach.
}
The star product of the regulated star-algebra projector
with itself approximately reproduces itself.
In the process of gluing, the midpoint of the boundary
of each projector, where an operator is inserted, is mapped to
a point which is very close to the midpoint of the boundary
of the glued surface state.
Therefore, the two operators from each of the projector
can be replaced by the leading term of their operator product
expansion (OPE).
On the other hand, the only effect of the BRST operator $Q_B$
acting on the regulated projector with an operator insertion
is to make the BRST transformation of the inserted operator.
If we take the operator to be a $c$ ghost,
the leading term of the OPE is given by $c \partial c$,
\begin{equation}
  c(z) c(w) = - (z-w) c \partial c (w) + O \left( (z-w)^2 \right),
\end{equation}
and the BRST transformation of the $c$ ghost is also
$c \partial c$:
\begin{equation}
  Q_B \cdot c(w) = \oint \frac{dz}{2 \pi i} \, j_B (z) c(w)
  = c \partial c (w),
\end{equation}
where $j_B (z)$ is the BRST current.
Therefore, the equation of motion of Witten's string field theory,
$Q_B \ket{\Psi} + \ket{ \Psi \ast \Psi } = 0$,
can be satisfied at the leading order
since the two terms are the same surface state
with $c \partial c$ inserted at the midpoint of the boundary.

Similarly, it can also be easily seen how our ansatz solves
the equation of motion of vacuum string field theory
at the leading order.
When we contract a regulated star-algebra projector with
a state in the Fock state, the open-string midpoint
approaches the midpoint of the boundary
\cite{Rastelli:2001vb, Gaiotto:2002kf}.
Therefore, the two $c$ ghosts,
one from the kinetic operator ${\cal Q}$ inserted
at the open-string midpoint and the other one
inserted at the midpoint of the boundary,
can be replaced by the leading term of the OPE,
which is again given by $c \partial c$.

The organization of the rest of the paper is as follows.
In Section 2,
the leading-order solution $| \Psi^{(0)} \rangle$
is constructed and its properties are studied.
In Section 3, it is shown that the leading-order solution
with an opposite sign, $- | \Psi^{(0)} \rangle$, also solves
the equation of motion of vacuum string field theory
at the leading order.
In Section 4,
the solutions $| \Psi^{(2)}_1 \rangle$ and $| \Psi^{(2)}_2 \rangle$
are constructed by taking into account next-to-leading terms,
and the energy density is computed for these solutions.
Section 5 is devoted to discussion.
Our conventions and terminology
on the conformal field theory (CFT) formulation
of string field theory \cite{LeClair:1988sp, LeClair:1988sj}
are summarized in Appendix A.
An explicit expression of the quantity
$\langle \Psi^{(2)} | \Psi^{(2)} \ast \Psi^{(2)} \rangle$
for an arbitrary $t$ is presented in Appendix B.

\section{Solution at the leading order}
\setcounter{equation}{0}

\subsection{Regulated butterfly state}

The regulated butterfly state $\ket{B_t}$
labeled by $t$ in the range $0 \le t < 1$
\cite{Schnabl:2002ff, Gaiotto:2002kf}
is defined by
\begin{equation}
  \vev{ \phi | B_t } = \vev{ f_t \circ \phi (0) }
\end{equation}
for any state $\ket{\phi}$ in the Fock space, where
\begin{equation}
  f_t (\xi) = \frac{\xi}{\sqrt{1 + t^2 \xi^2}}.
\end{equation}
All CFT correlation functions in this paper are evaluated
on an upper-half plane, and we use the doubling trick.
The butterfly state $\ket{B}$ is given
by the regulated butterfly state in the limit $t \to 1$.
It is a singular state like other star-algebra projectors
such as the sliver state \cite{Moore:2001fg}.
The singularity can be seen, for example, by the fact that
the open-string midpoint $f_t (i)$ reaches the boundary
in the limit $t \to 1$.
However, an inner product of the regulated butterfly state
with a state $\ket{\phi}$ in the Fock space is well defined
even in the limit $t \to 1$ and is given by
\begin{equation}
  \vev{ \phi | B } = \lim_{t \to 1} \vev{ \phi | B_t }
                   = \vev{ f_B \circ \phi (0) }
\end{equation}
where
\begin{equation}
  f_B (\xi) = \frac{\xi}{\sqrt{1 + \xi^2}}.
\end{equation}
The regulated butterfly state reduces to
the $SL(2,R)$-invariant vacuum $\ket{0}$ when $t=0$.

We can use different conformal transformations
to represent the same surface state.
For example, the regulated butterfly state
can also be represented as
\begin{equation}
  \vev{ \phi | B_t } = \vev{ h_t \circ \phi (0) }
\end{equation}
for any state $\ket{\phi}$ in the Fock space,
where $h_t (\xi)$ is the following conformal transformation
with a parameter $p$:
\begin{equation}
  h_t (\xi) = \frac{\xi}{\xi + p \sqrt{1 + t^2 \xi^2}}.
\label{h_t}
\end{equation}
The conformal transformation $h_t (\xi)$ is related to
$f_t (\xi)$ by an $SL(2,R)$ transformation $z/(z+p)$ 
which maps the infinity to $1$.
The conformal transformations
with different values of $p$ for (\ref{h_t})
are all equivalent and define the same state.
This representation will be useful when there is an operator
insertion at the infinity
in the representation in terms of $f_t (\xi)$.
In this representation, the inner product $\vev{ \phi | B }$
in the butterfly limit is given by
\begin{equation}
  \vev{ \phi | B } = \vev{ h_B \circ \phi (0) },
\end{equation}
where
\begin{equation}
  h_B (\xi) = \frac{\xi}{\xi + p \sqrt{1 + \xi^2}}.
\end{equation}

Another useful representation of the regulated butterfly state
$\ket{B_t}$ is given by making an inversion $I(z)=-1/z$
to the conformal transformation $f_t (\xi)$:
\begin{equation}
  I \circ f_t (\xi) = -\frac{\sqrt{1 + t^2 \xi^2}}{\xi}.
\end{equation}
Using this conformal transformation,
the regulated butterfly state is represented as
\begin{equation}
  \vev{ \phi | B_t }
  = \vev{ I \circ f_t \circ \phi (0) }.
\label{B_t-another}
\end{equation}
The coordinate $z'=I \circ f_t (\xi)$ is related to
the coordinate $z=h_t (\xi)$ by
\begin{equation}
  z' = \frac{z-1}{p z}, \quad z = \frac{1}{1 - p z'}.
\end{equation}

Finally, the regulated butterfly state has
a simple representation
in the operator formalism
\cite{Gross:1986ia, Gross:1986fk, Ohta:wn, Cremmer:if, Samuel:1986wp}
in terms of a single Virasoro generator
\cite{Schnabl:2002ff, Gaiotto:2002kf}.
It is given by
\begin{equation}
  \ket{B_t} = \exp \left( -\frac{t^2}{2} L_{-2} \right) \ket{0}.
\end{equation}

The conformal transformation associated with
the star product of the regulated butterfly state
$\ket{B_t \ast B_t}$ was derived by Schnabl \cite{Schnabl:2002ff}.
After an appropriate rescaling of Schnabl's expression
discussed in \cite{Okawa:2003cm},
the state $\ket{B_t \ast B_t}$ is represented as
\begin{equation}
  \vev{ \phi | B_t \ast B_t }
  = \vev{ \widetilde{f}_t \circ \phi (0) },
\end{equation}
where
\begin{equation}
  \widetilde{f}_t (\xi)
  = \sqrt{ \frac{3}{4} \frac{9-a^2}{(1-a^2)(a^2+3)}
    \left[ \tan^2 \left( \frac{2}{3}
    \arctan \sqrt{\frac{\xi^2 + t^2}{1 + t^2 \xi^2}} \right)
    - \frac{a^2}{3} \right]}
\label{f-tilde}
\end{equation}
with
\begin{equation}
  a \equiv \sqrt{3} \tan \left( \frac{2}{3} \arctan t \right).
\end{equation}
It can be easily verified that $\widetilde{f}_t (\xi)$ reduces to
$f_B (\xi)$ in the limit $t \to 1$:
\begin{equation}
  \widetilde{f}_t (\xi) = f_B (\xi) + O(1-t).
\end{equation}
Therefore,
\begin{equation}
  \vev{ \phi | B_t \ast B_t }
  = \vev{ \phi | B_t } + O(1-t).
\end{equation}
This shows that the butterfly state $\ket{B}$
is a star-algebra projector.
The relation between
the coordinate $z = \widetilde{f}_t (\xi)$
of $\vev{ \phi | B_t \ast B_t }$
and the coordinate $z' = I \circ f_t (\xi)$
of $\vev{ \phi | B_t }$ before gluing was derived
in \cite{Okawa:2003cm}. It is given by
\begin{equation}
  z'^2 
  = \frac{(1+t^2)(1+d^2 z^2)^{\frac{3}{2}}
          +(1-t^2)(1-\frac{4 \beta^2}{d^2} z^2)}
    {(1+d^2 z^2)^{\frac{3}{2}}-(1-\frac{4 \beta^2}{d^2} z^2)},
\label{z-z'}
\end{equation}
where
\begin{equation}
  \beta = \frac{2}{9-a^2} \sqrt{(1-a^2)(a^2+3)}, \quad
  d = 2 \sqrt{\frac{1-a^2}{9-a^2}}.
\end{equation}
The midpoint of the boundary $z'=0$ is mapped
to $z=\pm 1/\beta$, and the open-string midpoint $z'=i$
is mapped to $z=i/d$ \cite{Gaiotto:2002kf, Okawa:2003cm}.

\subsection{Solving the equation of motion at the leading order}

Our ansatz for a solution to Witten's string field theory
takes the form of the regulated butterfly state
with an operator insertion at the midpoint of the boundary.
Let us define $\ket{B_t ({\cal O})}$ by
\begin{equation}
  \vev{ \phi | B_t ({\cal O}) }
  = \vev{ I \circ f_t \circ \phi (0) \, {\cal O}(0) }
\end{equation}
for any state $\ket{\phi}$ in the Fock space.
As in the case of the regulated butterfly state $\ket{B_t}$,
the state $\ket{B_t ({\cal O})}$ itself becomes
singular in the limit $t \to 1$,
but its inner product with a state $\ket{\phi}$ in the Fock space
has a finite limit,
which we denote by $\vev{ \phi | B ({\cal O}) }$:
\begin{equation}
  \vev{ \phi | B ({\cal O}) }
  = \lim_{t \to1 }\vev{ \phi | B_t ({\cal O}) }
  = \vev{ I \circ f_B \circ \phi (0) \, {\cal O}(0) }.
\end{equation}
When the operator ${\cal O}$ is a $c$ ghost,
the representation of the state $\ket{B_t (c)}$
using $h_t (\xi)$ is given by
\begin{eqnarray}
  \vev{ \phi | B_t (c) }
  = \vev{ I \circ f_t \circ \phi (0) \, c(0) }
  = \frac{1}{p} \vev{ h_t \circ \phi (0) \, c(1) }
\end{eqnarray}
for any state $\ket{\phi}$ in the Fock space.
As we mentioned in the Introduction,
the state $\ket{B_t (c)}$ with an appropriate normalization
solves the equation of motion of Witten's string field theory
up to $O( \sqrt{1-t} )$.
In what follows we will use $\ket{\phi}$ to denote
a state in the Fock space.

Let us compute $\vev{ \phi | Q_B | B_t (c) }$
and $\vev{ \phi | B_t (c) \ast B_t (c) }$.
The quantity $\vev{ \phi | Q_B | B_t (c) }$ is given by
\begin{equation}
  \vev{ \phi | Q_B | B_t (c) }
  = \frac{1}{p} \vev{ h_t \circ \phi (0) \, Q_B \cdot c(1) },
\end{equation}
where $Q_B \cdot {\cal O} (w)$ denotes the BRST transformation
of the operator ${\cal O} (w)$ defined by
\begin{equation}
  Q_B \cdot {\cal O} (w)
  = \oint \frac{dz}{2 \pi i} \, j_B (z) {\cal O} (w),
\end{equation}
where the contour of the integral encircles $w$ counterclockwise.
The BRST current $j_B (z)$ here is defined by
\begin{equation}
  j_B (z) = c T^m (z) + \no{ bc \partial c } (z)
            + \frac{3}{2} \partial^2 c (z),
\end{equation}
where $T^m (z)$ denotes the energy-momentum tensor
of the matter sector and $\no{{\cal O}} (z)$ denotes
the normal ordering of the operator ${\cal O} (z)$.
Since the OPE between $j_B$ and $c$ is
\begin{equation}
  j_B (z) c(w) \sim \frac{1}{z-w} c \partial c (w),
\end{equation}
the BRST transformation of $c(w)$ is given by
\begin{equation}
  Q_B \cdot c(w) = c \partial c (w).
\end{equation}
Therefore, we find
\begin{equation}
  \vev{ \phi | Q_B | B_t (c) }
  = \frac{1}{p} \vev{ h_t \circ \phi (0) \, c \partial c(1) },
\label{Q_B-B_t-c}
\end{equation}
or, in other words,
\begin{equation}
  \vev{ \phi | Q_B | B_t (c) }
  = \vev{ \phi | B_t (c \partial c) }.
\end{equation}
As is clear from this derivation, $Q_B \ket{ B_t ({\cal O}) }$
is in general given by
\begin{equation}
  Q_B \ket{ B_t ({\cal O}) }
  = \ket{ B_t (Q_B \cdot {\cal O}) }.
\end{equation}

The only $t$ dependence of (\ref{Q_B-B_t-c}) is coming from
that of the conformal transformation $h_t (\xi)$.
Since
\begin{equation}
  h_t (\xi) = h_B (\xi) + O(1-t),
\end{equation}
the leading term of the expansion in $1-t$ is given by
\begin{equation}
  \vev{ \phi | Q_B | B_t (c) }
  = \frac{1}{p} \vev{ h_B \circ \phi (0) \, c \partial c(1) }
    + O(1-t),
\end{equation}
or
\begin{equation}
  \vev{ \phi | Q_B | B_t (c) }
  = \vev{ \phi | B(c \partial c) } + O(1-t).
\end{equation}

In order to compute the other quantity
$\vev{ \phi | B_t (c) \ast B_t (c) }$, we need to know
how the $c$ ghosts are mapped to the glued surface.
In the coordinate $z' = I \circ f_t (\xi)$
representing $\vev{ \phi | B_t (c) }$,
the $c$ ghost is inserted at $z'=0$.
This point is mapped to $z=\pm 1/\beta$
in the coordinate $z = \widetilde{f}_t (\xi)$
of $\vev{ \phi | B_t \ast B_t }$.
The derivatives $dz'/dz$ at these points were computed from
(\ref{z-z'}) in \cite{Okawa:2003cm} and are given by
\begin{equation}
  \left. \frac{dz'}{dz} \right|_{z=\pm \frac{1}{\beta}}
  = \sqrt{\frac{(1-t^4)(1-a^2)(3+a^2)}{48}}.
\end{equation}
Therefore, the $c$ ghost is mapped from the $z'$ coordinate
to the $z$ coordinate as follows:
\begin{equation}
  c(0) \to \sqrt{\frac{(1-t^4)(1-a^2)(3+a^2)}{48}} \,
           c \left( \pm \frac{1}{\beta} \right).
\end{equation}
Since
\begin{equation}
  \beta = \frac{\sqrt{2}}{3^{\frac{3}{4}}} \sqrt{1-t}
          +O \left( (1-t)^{\frac{3}{2}} \right),
\end{equation}
the two $c$ ghosts are sent to the infinity in the limit $t \to 1$.
It is therefore convenient to make an $SL(2,R)$ transformation
to bring the infinity to a finite point
to study the limit $t \to 1$.
If we make the conformal map $z/(z+p)$,
the $c$ ghosts are further transformed as
\begin{equation}
  c \left( \pm \frac{1}{\beta} \right)
  \to \frac{(1 \pm  p \beta)^2}{p \beta^2} \,
  c \left( \frac{1}{1 \pm p \beta} \right).
\end{equation}
Therefore, the quantity $\vev{ \phi | B_t (c) \ast B_t (c) }$
is given by
\begin{eqnarray}
  \vev{ \phi | B_t (c) \ast B_t (c) }
  &=& \frac{(1-t^4)(1-a^2)(3+a^2)}{48}
    \frac{(1 - p^2 \beta^2)^2}{p^2 \beta^4}
\nonumber \\
  && \times \vev{ \widetilde{h}_t \circ \phi (0) \,
            c \left( \frac{1}{1 + p \beta} \right)
            c \left( \frac{1}{1 - p \beta} \right) },
\end{eqnarray}
where
\begin{equation}
  \widetilde{h}_t (\xi)
  = \frac{\widetilde{f}_t (\xi)}{\widetilde{f}_t (\xi) + p}.
\end{equation}
In the limit $t \to 1$, the two $c$ ghosts approach
the point $z=1$ so that they can be replaced by the leading
term of their OPE, which is given by
\begin{equation}
  c \left( \frac{1}{1 + p \beta} \right)
  c \left( \frac{1}{1 - p \beta} \right)
  = 2 p \beta \, c \partial c (1) + O(\beta^3).
\end{equation}
Note that terms of $O(\beta^2)$ cancel.
The leading term of $\vev{ \phi | B_t (c) \ast B_t (c) }$
in the limit $t \to 1$ is given by
\begin{eqnarray}
  \vev{ \phi | B_t (c) \ast B_t (c) }
  &=& \frac{4 \sqrt{2}}{3^{\frac{1}{4}}} \frac{\sqrt{1-t}}{p}
    \vev{ h_B \circ \phi (0) \, c \partial c (1) }
    + O \left( (1-t)^{\frac{3}{2}} \right)
\nonumber \\
  &=& \frac{4 \sqrt{2}}{3^{\frac{1}{4}}} \sqrt{1-t}
    \vev{ \phi | B(c \partial c) }
    + O \left( (1-t)^{\frac{3}{2}} \right).
\label{phi-B_t-c^2}
\end{eqnarray}
Therefore, the leading term of $\vev{ \phi | B_t (c) \ast B_t (c) }$
is proportional to that of $\vev{ \phi | Q_B | B_t (c) }$.
If we define
\begin{equation}
  | \Psi^{(0)} \rangle = -\frac{3^{\frac{1}{4}}}{4 \sqrt{2}}
                          \frac{1}{\sqrt{1-t}} \ket{ B_t (c) },
\label{Psi^0}
\end{equation}
then the state $| \Psi^{(0)} \rangle$ solves the equation of motion
of Witten's string field theory up to $O(\sqrt{1-t})$:
\begin{equation}
  \langle \phi | Q_B | \Psi^{(0)} \rangle
  + \langle \phi | \Psi^{(0)} \ast \Psi^{(0)} \rangle
  = O(\sqrt{1-t})
\end{equation}
for any state $\ket{\phi}$ in the Fock space.
If we take the limit $t \to 1$,
the state $| \Psi^{(0)} \rangle$
formally solves the equation of motion exactly.
However, the regulated butterfly state $\ket{B_t}$ becomes
singular and the coefficient diverges as $1/\sqrt{1-t}$
in the limit so that
we do not intend to take the strict $t \to 1$ limit.
On the other hand, the state $| \Psi^{(0)} \rangle$
is well defined as long as $1-t$ is finite.
If we choose $t$ to be 0.9999, for example,
the equation of motion is solved with fairly good precision
for any state $\ket{\phi}$ in the Fock space.

\subsection{Energy density at the leading order}

Let us next evaluate the energy density of the solution.
We need to compute the inner products
$\vev{ B_t (c) | Q_B | B_t (c) }$
and $\vev{ B_t (c) | B_t (c) \ast B_t (c) }$, but
similar computations have been done in \cite{Okawa:2003cm}
so that we can make use of the results in \cite{Okawa:2003cm}.

When we deal with the star multiplication
of the regulated butterfly state,
it is convenient to use the $\hat{z}$ coordinate
\cite{Gaiotto:2002kf} defined by
\begin{equation}
  \hat{z} = \arctan \xi.
\end{equation}
In the $\hat{z}$ coordinate,
either of the left and right halves of the open string
of the regulated butterfly state
is mapped to a semi-infinite line
parallel to the imaginary axis in the upper-half plane
so that gluing can be made simply by translation.
We glue two or three regulated butterfly states
for $\vev{ B_t (c) | Q_B | B_t (c) }$
or $\vev{ B_t (c) | B_t (c) \ast B_t (c) }$, respectively,
in this way.
We then map the resulting surface to an upper-half plane
by a conformal transformation.
Let us denote the coordinate of the upper-half plane by $z$.
The relation between the $z$ coordinate and the coordinate
$z' = I \circ f_t (\xi)$
before gluing was derived in \cite{Okawa:2003cm}.
In the case of $\vev{ B_t (c) | Q_B | B_t (c) }$,
it is given by
\begin{equation}
  \frac{4 (z'^2-t^2)}{\left[ z'^2-(1+t^2) \right]^2}
  = \frac{(1-z^2)^2-4 q^2 z^2}{4 (1+q^2) z^2},
\label{z-z'-2}
\end{equation}
where
\begin{equation}
  q = \frac{2 t}{1-t^2},
\label{q}
\end{equation}
and the relation for $\vev{ B_t (c) | B_t (c) \ast B_t (c) }$
is given as follows:
\begin{equation}
  \frac{4 (z'^2-t^2)}{\left[ z'^2-(1+t^2) \right]^2}
  = \frac{z^2 (z^2-3)^2-q^2 (1-3 z^2)^2}{(1+q^2)(1-3 z^2)^2}.
\label{z-z'-3}
\end{equation}
The $c$ ghost at $z'=0$ is mapped to
\begin{equation}
  c(0) \to \frac{\sqrt{1-t^4}}{2} \, c(\pm 1)
\end{equation}
in the $z$ coordinate of $\vev{ B_t (c) | Q_B | B_t (c) }$,
and to
\begin{equation}
  c(0) \to \frac{3}{8} \sqrt{1-t^4} \, c(-\sqrt{3}), \quad
  \frac{3}{2} \sqrt{1-t^4} \, c(0), \quad
  \frac{3}{8} \sqrt{1-t^4} \, c(\sqrt{3})
\end{equation}
in the $z$ coordinate
of $\vev{ B_t (c) | B_t (c) \ast B_t (c) }$.
The inner product $\vev{ B_t (c) | Q_B | B_t (c) }$
is therefore given by
\begin{eqnarray}
  \vev{ B_t (c) | Q_B | B_t (c) }_{density}
  &=& \frac{1-t^4}{4} \vev{ c(-1) \, Q_B \cdot c(1) }_{density}
\nonumber \\
  &=& \frac{1-t^4}{4} \vev{ c(-1) \, c \partial c (1) }_{density}
  = -(1-t^4),
\end{eqnarray}
where the subscript $density$ denotes that
the quantity is divided by the volume factor of space-time.
We use this notation for both inner products of string fields
and CFT correlation functions.
Similarly, $\vev{ B_t (c) | B_t (c) \ast B_t (c) }$ is computed
as follows:
\begin{eqnarray}
  \vev{ B_t (c) | B_t (c) \ast B_t (c) }_{density}
  &=& \frac{27}{128} (1-t^4)^{\frac{3}{2}}
    \vev{ c(-\sqrt{3}) \, c(0) \, c(\sqrt{3}) }_{density}
\nonumber \\
  &=& - \left( \frac{3 \sqrt{3}}{4} \right)^3
        (1-t^4)^{\frac{3}{2}}.
\label{B_t-c^3}
\end{eqnarray}

In both cases, the inner product cancels the singularity
coming from the normalization factor so that
$\langle \Psi^{(0)} | Q_B | \Psi^{(0)} \rangle$
and $\langle \Psi^{(0)} | \Psi^{(0)} \ast \Psi^{(0)} \rangle$
have a finite density in the limit $t \to 1$.
Since $\langle \Psi^{(0)} | Q_B | \Psi^{(0)} \rangle$
and $\langle \Psi^{(0)} | \Psi^{(0)} \ast \Psi^{(0)} \rangle$
are well defined, we can discuss
the equation of motion contracted with the solution itself.
Let us compute the ratio ${\cal R} [ \Psi^{(0)} ]$
defined in (\ref{R-definition}) in the limit $t \to 1$:
\begin{equation}
  \lim_{t \to 1} {\cal R} [ \Psi^{(0)} ]
  = \lim_{t \to 1}
  \frac{\langle \Psi^{(0)} | Q_B | \Psi^{(0)} \rangle}
  {\langle \Psi^{(0)} | \Psi^{(0)} \ast \Psi^{(0)} \rangle}
  = -2^{\frac{15}{2}} \cdot 3^{-\frac{19}{4}} \simeq -0.9804.
\label{Psi^0-ratio}
\end{equation}
The ratio is not exactly $-1$, but rather close to $-1$
so that the equation of motion is satisfied
with fairly good precision
even when it is contracted with the solution itself.
In other words, the solution $| \Psi^{(0)} \rangle$ satisfies
the condition (ii-a) in the Introduction with good precision.

Let us next evaluate the energy density of the solution
${\cal E}$ defined by
\begin{equation}
  {\cal E} [ \Psi ]
  = \frac{1}{\alpha'^3 g_T^2} \left[
  \frac{1}{2} \vev{ \Psi | Q_B | \Psi }_{density}
  + \frac{1}{3} \vev{ \Psi | \Psi \ast \Psi }_{density} \right].
\end{equation}
The energy density of the solution ${\cal E} [ \Psi^{(0)} ]$
divided by the D25-brane tension $T_{25}$
in the limit $t \to 1$ is given by
\begin{equation}
  \lim_{t \to 1} \frac{{\cal E} [ \Psi^{(0)} ]}{T_{25}}
  = 2 \pi^2 (-2^{-4} \cdot 3^{\frac{1}{2}}
             +2^{-\frac{21}{2}} \cdot 3^{\frac{17}{4}})
    \simeq -0.6838.
\end{equation}
Therefore, the absolute value of the energy density
for the solution $| \Psi^{(0)} \rangle$ gives
approximately 68\% of the D25-brane tension.
This value is very close to that obtained by level truncation
at level 0 \cite{Sen:1999nx},
${\cal E}/T_{25} = -4096 \pi^2/59049 \simeq -0.6846$,
but not exactly the same.
We conclude that our solution $| \Psi^{(0)} \rangle$
at the leading order is as good as the level-0 solution
by level truncation
concerning the condition (ii-b) in the Introduction.

We have seen that our solution $| \Psi^{(0)} \rangle$
approximately satisfies the condition (ii-a),
but there are no a priori reasons to expect
that this should be the case.
As can be seen from the expression (\ref{B_t-c^3}),
the $c$ ghosts can no longer be replaced by $c \partial c$
so that the mechanism responsible for the condition (i)
does not work when the equation of motion is contracted
with the solution itself.
There are also no a priori reasons to expect that
the energy density of our solution $| \Psi^{(0)} \rangle$
is of the same order as the D25-brane tension
just like there are no a priori reasons to expect
that the ordinary level-truncation approximation should work.

\subsection{Comparison with a solution by a variational method}

We can perform an analysis similar to level truncation
by truncating the string field to the single mode $\ket{B_t (c)}$,
\begin{equation}
  \ket{ \Psi } = \frac{\tilde{x}}{\sqrt{1-t^4}} \ket{B_t (c)},
\end{equation}
and by taking a variation with respect to the parameter $\tilde{x}$.
It reduces to the ordinary level-truncation analysis at level 0
when we set $t=0$.
The energy density divided by the D25-brane tension is
given by
\begin{equation}
  \frac{{\cal E} (\tilde{x})}{T_{25}}
  = 2 \pi^2 \left[ -\frac{1}{2} \tilde{x}^2
    -\frac{1}{3} \left( \frac{3 \sqrt{3}}{4} \right)^3 \tilde{x}^3
    \right].
\end{equation}
This is independent of $t$
and is exactly the same as in the case of the ordinary
truncation to $\ket{\Psi} = \tilde{x} \, c_1 \ket{0}$.
Therefore, the critical value $\tilde{x}_c$ determined
by the condition
\begin{equation}
  \left. \frac{d}{d \tilde{x}} \, {\cal E} (\tilde{x})
  \right|_{\tilde{x}=\tilde{x}_c} = 0
\end{equation}
and ${\cal E} (\tilde{x_c})/T_{25}$ are also the same
as in the case of the ordinary level truncation:
\begin{equation}
  \tilde{x}_c = -\left( \frac{4}{3 \sqrt{3}} \right)^3, \quad
  \frac{{\cal E} (\tilde{x}_c)}{T_{25}}
  = -\frac{4096 \pi^2}{59049} \simeq -0.6846.
\end{equation}
The string field with $\tilde{x}=\tilde{x}_c$,
\begin{equation}
  | \widetilde{\Psi}^{(0)} \rangle
  = -\left( \frac{4}{3 \sqrt{3}} \right)^3
    \frac{1}{\sqrt{1-t^4}} \ket{B_t (c)},
\end{equation}
by construction satisfies the condition (ii-a),
\begin{equation}
  \langle \widetilde{\Psi}^{(0)} |
  Q_B | \widetilde{\Psi}^{(0)} \rangle
  +   \langle \widetilde{\Psi}^{(0)} |
  \widetilde{\Psi}^{(0)} \ast \widetilde{\Psi}^{(0)} \rangle = 0,
\end{equation}
but does not solve the equation of motion when it is contracted
with a generic state in the Fock space.
The normalization of our solution $| \Psi^{(0)} \rangle$ is in fact
numerically close to that of $| \widetilde{\Psi}^{(0)} \rangle$
in the limit $t \to 1$:
\begin{equation}
  \lim_{t \to 1}
  \frac{| \widetilde{\Psi}^{(0)} \rangle}{| \Psi^{(0)} \rangle}
  = 2^{\frac{15}{2}} \cdot 3^{-\frac{19}{4}} \simeq 0.9804.
\end{equation}
This is why our solution $| \Psi^{(0)} \rangle$ approximately
satisfies the equation 
$\langle \Psi^{(0)} | Q_B | \Psi^{(0)} \rangle
+ \langle \Psi^{(0)} | \Psi^{(0)} \ast \Psi^{(0)} \rangle = 0$,
and approximately reproduces the value of the energy density
obtained by the ordinary level truncation at level 0,
but again we do not know why this is the case.

\section{Solution
to both Witten's and vacuum string field theories}
\setcounter{equation}{0}

If we expand the action of Witten's string field theory
(\ref{Witten-action}) around the solution $\ket{\Psi_0}$
corresponding to the tachyon vacuum,
the resulting action will take the same form
except for the kinetic operator:
\begin{equation}
  S = S_0 -\frac{1}{\alpha'^3 g_T^2} \left[
      \frac{1}{2} \vev{ \Psi | {\cal Q} | \Psi }
      + \frac{1}{3} \vev{ \Psi | \Psi \ast \Psi } \right],
\end{equation}
where $S_0$ is the value of the action for $\ket{\Psi_0}$,
and ${\cal Q}$ is given by
\begin{equation}
  {\cal Q} \ket{ \Psi } = Q_B \ket{ \Psi }
  + \ket{ \Psi_0 \ast \Psi }
  + \ket{ \Psi \ast \Psi_0 }.
\end{equation}
The equation of motion at the tachyon vacuum,
\begin{equation}
  {\cal Q} \ket{ \Psi } + \ket{ \Psi \ast \Psi } = 0,
\end{equation}
can be solved by $\ket{ \Psi_0 }$ with an opposite sign,
\begin{equation}
  - {\cal Q} \ket{\Psi_0}
  + \ket{ \Psi_0 \ast \Psi_0 } = 0,
\end{equation}
and it describes a D25-brane
as an excitation from the tachyon vacuum.

It was conjectured in \cite{Rastelli:2000hv}
that ${\cal Q}$ can be made purely
of ghost fields by field redefinition, and string field theory
with this conjectured form of the kinetic operator is called
vacuum string field theory
\cite{Rastelli:2000hv, Rastelli:2001jb, Rastelli:2001uv}.
A more specific conjecture on ${\cal Q}$ was put forward later
in \cite{Gaiotto:2002kf}.
The kinetic operator ${\cal Q}$ does not seem to be
made purely of ghost fields when we expand the action
around the approximate solution constructed by level truncation.
It was conjectured \cite{Gaiotto:2002kf}
that there exists a one-parameter family of field redefinition
which takes ${\cal Q}$ to the following form:
\begin{equation}
  {\cal Q} = \frac{Q}{\epsilon} \left[ 1 + o(\epsilon) \right],
\label{Q-conjecture}
\end{equation}
where $Q$ is a $c$-ghost insertion at the open-string midpoint,
\begin{equation}
  Q = \frac{1}{2i} (c(i) - c(-i)),
\label{Q}
\end{equation}
and $\epsilon$ corresponds to the parameter
of the field redefinition. We denoted
terms which vanishes in the limit $\epsilon \to 0$
by $o(\epsilon)$.
In the singular limit $\epsilon \to 0$, the midpoint $c$-ghost
insertion $Q$ dominates in the kinetic operator ${\cal Q}$
with an infinite coefficient.

Since the string field of vacuum string field theory
is related to that of Witten's string field theory
by field redefinition,
the tachyon vacuum solution of Witten's string field theory
with an opposite sign
does not necessarily solve the equation of motion
of vacuum string field theory.
Interestingly, however, our leading-order solution
with an opposite sign, $- | \Psi^{(0)} \rangle$,
does solve the equation of motion of vacuum string field theory
at the leading order without field redefinition.

Let us first compute the quantity $\vev{ \phi | Q | B_t (c) }$.
It is convenient to rewrite it as
$\vev{ \phi | Q | B_t (c) } = \vev{ B_t (c) | Q | \phi }$.
Since the operator $Q$ is mapped to
\begin{eqnarray}
  && \frac{i}{2p} \sqrt{1-t^2} \left[
  (1-ip \sqrt{1-t^2})^2 \,
  c \left( \frac{1}{1-ip \sqrt{1-t^2}} \right) \right.
\nonumber \\
  && \qquad \qquad \qquad - (1+ip \sqrt{1-t^2})^2 \left.
  c \left( \frac{1}{1+ip \sqrt{1-t^2}} \right) \right]
\end{eqnarray}
by the conformal transformation $h_t (\xi)$,
the quantity $\vev{ B_t (c) | Q | \phi }$
is given by
\begin{eqnarray}
  && \vev{ B_t (c) | Q | \phi }
  = \frac{i}{2 p^2} \sqrt{1-t^2}
  \left[ (1-ip \sqrt{1-t^2})^2
  \vev{ c (1) \,
        c \left( \frac{1}{1-ip \sqrt{1-t^2}} \right)
        h_t \circ \phi (0) } \right.
\nonumber \\ && \qquad \qquad \qquad \qquad \qquad
  \left. - (1+ip \sqrt{1-t^2})^2
  \vev{ c (1) \,
        c \left( \frac{1}{1+ip \sqrt{1-t^2}} \right)
        h_t \circ \phi (0) } \right].
\nonumber \\
\end{eqnarray}
In the limit $t \to 1$,
the $c$ ghost coming from $Q$ approaches $c(1)$ so that
the two operators can be replaced by the leading term
of their OPE:
\begin{equation}
  c(1) \, c \left( \frac{1}{1 \pm ip \sqrt{1-t^2}} \right)
  = \mp ip \sqrt{2(1-t)} \, c \partial c (1) + O(1-t).
\end{equation}
The leading term of $\vev{ B_t (c) | Q | \phi }$
in the limit $t \to 1$ is given by
\begin{equation}
  \vev{ B_t (c) | Q | \phi }
  = -\frac{2}{p} (1-t)
  \vev{ c \partial c (1) \, h_B \circ \phi (0) }
  + O \left( (1-t)^2 \right).
\end{equation}
Note that terms of $O((1-t)^{3/2})$ cancel so that
the next-to-leading order is  $O((1-t)^2)$.
Since the leading term is proportional to
that of $\vev{ \phi | B_t (c) \ast B_t (c) }$
in (\ref{phi-B_t-c^2}),
our solution with an opposite sign,
$-| \Psi^{(0)} \rangle$, solves the equation of motion
of vacuum string field theory up to $O(\sqrt{1-t})$,
\begin{equation}
  - \langle \phi | {\cal Q} | \Psi^{(0)} \rangle
  + \langle \phi | \Psi^{(0)} \ast \Psi^{(0)} \rangle
  = O(\sqrt{1-t})
\end{equation}
for any state $\ket{\phi}$ in the Fock space
if the scaling between $\epsilon$ and $1-t$ is given by
\begin{equation}
  \epsilon = 2(1-t).
\label{epsilon}
\end{equation}

As we mentioned in the Introduction,
it is easily understood why $\ket{B_t (c)}$
with an appropriate normalization solves
the equation of motion of vacuum string field theory
at the leading order.
Both in $\vev{ \phi | Q | B_t (c) }$
and in $\vev{ \phi | B_t (c) \ast B_t (c) }$,
the two $c$ ghosts approach
the midpoint of the boundary of the surface
and can be replaced by
$c \partial c$ which is the leading term of the OPE.
The existence of our approximate solution
of Witten's string field theory seems to be consistent
with the vacuum string field theory conjecture at this order.
It would be interesting to explore the relation
between our approach and vacuum string field theory
in a more systematic way.

It was pointed out in \cite{Okawa:2003cm} that
subleading terms in (\ref{Q-conjecture}) are necessary
in order for vacuum string field theory
to have a parameter corresponding to
the string coupling constant.
While the leading term $Q/\epsilon$ dominates
in $\vev{ \phi_1 | {\cal Q} | \phi_2 }$
for any pair of states $\ket{\phi_1}$ and $\ket{\phi_2}$
in the Fock space,
the subleading terms may contribute at the same order
as the leading term in other quantities.
For example, $\vev{ \phi | Q | B_t (c) }/\epsilon$
and $\vev{ \phi | Q_B | B_t (c) }$ are the same order
when $\epsilon \sim 1-t$.
Therefore, the coefficient in (\ref{epsilon})
can be modified if we take into account the subleading terms.

Let us next consider
the equation of motion of vacuum string field theory
contracted with the solution itself.
The computation of $\vev{ B_t (c) | Q | B_t (c) }$
is almost parallel to that of
$\vev{ B_t (c) | Q_B | B_t (c) }$.
The operator $Q$ is mapped to
\begin{equation}
  -\frac{1-t^2}{2 (1+t^2)} \left( c(i)+c(-i) \right)
\end{equation}
in the $z$ coordinate we introduced for
the computation of $\vev{ B_t (c) | Q_B | B_t (c) }$
so that the density of $\vev{ B_t (c) | Q | B_t (c) }$ is given by
\begin{eqnarray}
  && \vev{ B_t (c) | Q | B_t (c) }_{density}
\nonumber \\
  && = -\frac{(1-t^2)^2}{8}
  \left[ \vev{ c(-1) \, c(i) \, c(1) }_{density}
  + \vev{ c(-1) \, c(-i) \, c(1) }_{density} \right]
\nonumber \\
  && = (1-t^2)^2.
\end{eqnarray}
Therefore, the quantity
$\langle \Psi^{(0)} | Q | \Psi^{(0)} \rangle /\epsilon$
has a finite density in the limit $t \to 1$
if $\epsilon$ scales as $1-t$.
When $\epsilon = 2(1-t)$, the ratio of
$\langle \Psi^{(0)} | Q | \Psi^{(0)} \rangle /\epsilon$
to $\langle \Psi^{(0)} | \Psi^{(0)} \ast \Psi^{(0)} \rangle$
in the limit $t \to 1$ is given by
\begin{equation}
  \lim_{t \to 1} \frac{1}{2 (1-t)}
  \frac{\langle \Psi^{(0)} | Q | \Psi^{(0)} \rangle}
  {\langle \Psi^{(0)} | \Psi^{(0)} \ast \Psi^{(0)} \rangle}
  = 2^{\frac{13}{2}} \cdot 3^{-\frac{19}{4}} \simeq 0.4902.
\end{equation}
The deviation of the value from $1$ is much worse than
the case of (\ref{Psi^0-ratio}).
In other words, the analogues of the conditions (i) and (ii-a)
for vacuum string field theory are not compatible
for the combination of $Q/\epsilon$
and $- | \Psi^{(0)} \rangle$ at the leading order.
This result is independent of the scaling among
$t$, $\epsilon$, and the normalization of the solution
in (\ref{Psi^0}).
To make this point clearer, let us compute
the following quantity introduced in \cite{Okawa:2003cm}:
\begin{equation}
  \frac{\vev{\Psi \ast \Psi | \phi}}{\vev{\Psi | {\cal Q} | \phi}}
  \frac{\vev{\Psi | {\cal Q} | \Psi}}{\vev{\Psi \ast \Psi | \Psi}}.
\end{equation}
If this quantity is different from $1$, the equation of motion
contracted with the solution itself is not compatible with
the one contracted with a state $\ket{\phi}$ in the Fock space.
This quantity is independent of the normalizations of
${\cal Q}$, $\ket{\Psi}$, and $\ket{\phi}$ so that
if ${\cal Q}$ is dominated by $Q/\epsilon$
and $\ket{\Psi}$ is dominated by $| \Psi^{(0)} \rangle$
in the limit $t \to 1$, it reduces to
\begin{equation}
  \lim_{t \to 1}
  \frac{\vev{B_t (c) \ast B_t (c) | \phi}}
       {\vev{B_t (c) | Q | \phi}}
  \frac{\vev{B_t (c) | Q | B_t (c)}}
       {\vev{B_t (c) \ast B_t (c) | B_t (c)}}
  = 2^{\frac{13}{2}} \cdot 3^{-\frac{19}{4}} \simeq 0.4902.
\label{imcompatibility-1}
\end{equation}
A similar result was derived for the twisted regulated
butterfly state $\ket{B'_t}$ in \cite{Okawa:2003cm}:
\begin{equation}
  \lim_{t \to 1}
  \frac{\vev{B'_t \ast B'_t | \phi}}{\vev{B'_t | Q | \phi}}
  \frac{\vev{B'_t | Q | B'_t}}{\vev{B'_t \ast B'_t | B'_t}}
  = \frac{\sqrt{2}}{3} \simeq 0.4714.
\label{imcompatibility-2}
\end{equation}
In both cases, the analogues of the conditions (i) and (ii-a)
for vacuum string field theory are not compatible
if we assume that $Q/\epsilon$ dominates in the kinetic operator
and the solution is dominated by either $\ket{B_t (c)}$
or $\ket{B'_t}$ with an appropriate normalization,
and this conclusion holds
whatever scaling we may take for $t$, $\epsilon$,
and the normalization of the solution.
As was demonstrated in \cite{Okawa:2003cm},
however, subleading terms of the kinetic operator ${\cal Q}$
can contribute  to the quantity $\vev{ \Psi | {\cal Q} | \Psi}$
at the same order as the leading term given by $Q/\epsilon$.
Subleading terms of the solution may also contribute
to $\vev{ \Psi | {\cal Q} | \Psi}$
and $\vev{ \Psi \ast \Psi | \Psi }$ at the same order.
In fact, as we will show in the next section,
the condition (ii-a) can be satisfied with better precision
when we incorporate next-to-leading terms of the solution
in the case of Witten's string field theory.
Contributions from subleading terms in ${\cal Q}$ and
in the solution may provide a resolution
of the incompatibility in (\ref{imcompatibility-1})
or in (\ref{imcompatibility-2}).
However, the relevance of the subleading terms in ${\cal Q}$
may ruin the factorization of the matter and ghost sectors
at the leading term.
We thus recognize that this is an important issue
for the vacuum string field theory conjecture
to be studied further.

\section{Solutions at the next-to-leading order}
\setcounter{equation}{0}

\subsection{Solving the equation of motion
at the next-to-leading order}

The solution $| \Psi^{(0)} \rangle$ in Section 2 solves
the equation of motion of Witten's string field theory
up to $O(\sqrt{1-t})$:
\begin{equation}
  \langle \phi | Q_B | \Psi^{(0)} \rangle
  + \langle \phi | \Psi^{(0)} \ast \Psi^{(0)} \rangle
  = O(\sqrt{1-t}).
\end{equation}
Let us try to improve the solution such that
the equation is satisfied up to a higher power of $1-t$.

There are two important sources of contributions
at the next-to-leading order.
First, there are contributions from
the next-to-leading terms
of the OPE in $\vev{ \phi | B_t (c) \ast B_t (c) }$.
Since the OPE of $c(-z)$ and $c(z)$ is given by
\begin{equation}
  c(-z) c(z) = 2 z \, c \partial c (0)
  + \frac{1}{3} z^3 \, c \partial^3 c (0)
  - z^3 \, \partial c \partial^2 c (0) + O(z^5),
\end{equation}
the inner products $\langle \phi | B(c \partial^3 c) \rangle$
and $\langle \phi | B(\partial c \partial^2 c) \rangle$
will appear in $\vev{ \phi | B_t (c) \ast B_t (c) }$
at the next-to-leading order.

Second, while the conformal transformations $f_t (\xi)$
and $\widetilde{f}_t (\xi)$ coincide at the leading order,
they differ at the next-to-leading order:
\begin{eqnarray}
  f_t (\xi) &=& \frac{\xi}{\sqrt{1 + t^2 \xi^2}}
  = \frac{\xi}{\sqrt{1 + \xi^2}}
    + (1-t) \left( \frac{\xi}{\sqrt{1 + \xi^2}} \right)^3
    + O \left( (1-t)^2 \right)
\nonumber \\
  &=& f_B (\xi) + (1-t) f_B (\xi)^3 + O \left( (1-t)^2 \right),
\\
  \widetilde{f}_t (\xi)
  &=& f_B (\xi) + \frac{1}{\sqrt{3}} (1-t) f_B (\xi)^3
    + O \left( (1-t)^2 \right).
\end{eqnarray}
The next-to-leading terms are proportional
to $f_B (\xi)^3$ in both cases,
but the coefficients are different.
Up to this order,
the conformal transformation $z = f_t (\xi)$ can be regarded as
a combination of two transformations given by
$\tilde{z} = f_B (\xi)$ and an infinitesimal one
$z = \tilde{z} + (1-t) \tilde{z}^3 + O((1-t)^2)$.
Therefore, $f_t \circ \phi (0)$ can be represented
in the expansion with respect to $1-t$ as follows:
\begin{equation}
  f_t \circ \phi (0) = f_B \circ \phi (0)
  + (1-t) \oint \frac{dz}{2 \pi i} \, z^3 T(z) f_B \circ \phi (0)
  + O \left( (1-t)^2 \right),
\end{equation}
where the contour encircles the point $f_B \circ \phi (0)$
counterclockwise.
Similarly, $\widetilde{f}_t \circ \phi (0)$ is given by
\begin{equation}
  \widetilde{f}_t \circ \phi (0) = f_B \circ \phi (0)
  + \frac{1}{\sqrt{3}} (1-t)
  \oint \frac{dz}{2 \pi i} \, z^3 T(z) f_B \circ \phi (0)
  + O \left( (1-t)^2 \right).
\end{equation}
As we have shown in Section 2, $\vev{ \phi | Q_B | B_t (c) }$
is equal to $\vev{ \phi | B_t (c \partial c) }$.
Let us explicitly evaluate it
up to the next-to-leading order in $1-t$. Since
\begin{equation}
  I \circ f_t \circ \phi (0) = I \circ f_B \circ \phi (0)
  - (1-t) \oint \frac{dz}{2 \pi i} \frac{1}{z} \, T(z) \,
  I \circ f_B \circ \phi (0)
  + O \left( (1-t)^2 \right),
\end{equation}
where the contour encircles the point $I \circ f_B (0) = \infty$
counterclockwise, the quantity $\vev{ \phi | B_t (c \partial c) }$
is given by
\begin{eqnarray}
  \vev{ \phi | B_t (c \partial c) }
  &=& \vev{ I \circ f_t \circ \phi (0) \, c \partial c (0) }
\nonumber \\
  &=& \vev{ I \circ f_B \circ \phi (0) \, c \partial c (0) }
    + (1-t) \vev{ I \circ f_B \circ \phi (0)
    \oint \frac{dz}{2 \pi i} \frac{1}{z} \, T(z) \,
    c \partial c (0) }
\nonumber \\
  && {} + O \left( (1-t)^2 \right),
\end{eqnarray}
where the contour encircles the origin counterclockwise.
We have to compute the first regular term
in the OPE of $T(z)$ and $c \partial c (0)$.
The energy-momentum tensor $T(z)$ is given by
\begin{equation}
  T(z) = T^m (z) + \no{ ( \partial b ) c } (z)
         -2 \, \partial ( \no{ bc } ) (z),
\end{equation}
and its OPE with $c \partial c (w)$ is
\begin{eqnarray}
  T(z) c \partial c (w)
  &=& -\frac{1}{(z-w)^2} \, c \partial c (w)
      +\frac{1}{z-w} \, c \partial^2 c (w)
\nonumber \\ &&
    {} + \frac{2}{3} \, c \partial^3 c (w)
    - \frac{3}{2} \, \partial c \partial^2 c (w)
    + c \partial c T^m (w) + O(z-w).
\end{eqnarray}
Therefore, $\vev{ \phi | B_t (c \partial c) }$ is given by
\begin{eqnarray}
  \vev{ \phi | B_t (c \partial c) }
  &=& \vev{ \phi | B (c \partial c) }
\nonumber \\ && {}
  + (1-t) \left[ \, \frac{2}{3} \langle \phi
          | B (c \partial^3 c) \rangle
  -\frac{3}{2} \langle \phi | B (\partial c \partial^2 c) \rangle
  + \vev{ \phi | B (c \partial c T^m) } \right]
\nonumber \\ && {}
  + O \left( (1-t)^2 \right).
\end{eqnarray}

To summarize, contributions from both sources consist of
$\langle \phi | B (c \partial^3 c) \rangle$,
$\langle \phi | B (\partial c \partial^2 c) \rangle$,
and $\vev{ \phi | B (c \partial c T^m) }$.
The mass dimension of the operators inserted in these states
is $1$, which is that of $c \partial c$ in the leading term
plus $2$.
In addition, $\vev{ \phi | B (c \partial c) }$ appearing
at the leading order will also appear
at the next-to-leading order.
In order to satisfy the equation of motion
up to the next-to-leading order,
let us incorporate a set of states $\ket{B_t ({\cal O})}$
with the mass dimension of ${\cal O}$ being $1$
into our ansatz.
There are three such operators which consist of
$b$, $c$, and $T^m$ and have the correct ghost number:
$| B_t(\partial^2 c) \rangle$,
$\ket{ B_t(c T^m) }$, and $\ket{ B_t(\no{ bc \partial c}) }$.
Our ansatz then takes the following form:
\begin{eqnarray}
  | \Psi^{(2)} \rangle &=& \frac{x}{\sqrt{1-t}} \ket{B_t(c)}
\nonumber \\
  && {} + \sqrt{1-t} \left[ u \, | B_t(\partial^2 c) \rangle
  + v \, \ket{ B_t(c T^m) }
  + w \, \ket{ B_t(\no{ bc \partial c}) } \right],
\end{eqnarray}
where the $t$ dependence in the coefficients has been chosen
for later convenience.
The four operators $c$, $\partial^2 c$, $c T^m$,
and $\no{ bc \partial c }$ correspond to
$c_1 \ket{0}$, $2 \, c_{-1} \ket{0}$, $L^m_{-2} c_1 \ket{0}$,
and $- b_{-2} c_0 c_1 \ket{0}$, respectively, which constitute
all the states up to level 2 in level truncation
without gauge fixing.

Let us compute $\langle \phi | Q_B | \Psi^{(2)} \rangle$
and $\langle \phi | \Psi^{(2)} \ast \Psi^{(2)} \rangle$.
Since
\begin{equation}
  \vev{ \phi | Q_B | B_t ({\cal O}) }
  = \vev{ \phi | B_t (Q_B \cdot {\cal O}) }
\end{equation}
for any operator ${\cal O}$,
we need to derive the BRST transformations of
$\partial^2 c$, $c T^m$, and $\no{ bc \partial c }$
for the computation of
$\langle \phi | Q_B | \Psi^{(2)} \rangle$.
The BRST transformation of $\partial^2 c$ can be easily
derived from that of $c$:
\begin{equation}
  Q_B \cdot \partial^2 c (w)
  = \partial^2 ( Q_B \cdot c (w) )
  = \partial c \partial^2 c (w) + c \partial^3 c (w).
\end{equation}
  From the OPE's,
\begin{eqnarray}
  j_B (z) \, c T^m (w)
  &\sim& -\frac{13}{(z-w)^3} \, c \partial c (w)
  -\frac{13}{2} \frac{1}{(z-w)^2} \, c \partial^2 c (w)
\nonumber \\ && {}
  -\frac{13}{6} \frac{1}{z-w} \, c \partial^3 c (w)
  -\frac{1}{z-w} \, c \partial c T^m (w),
\\
  j_B (z) \, \no{ bc \partial c } (w)
  &\sim& \frac{6}{(z-w)^3} \, c \partial c (w)
  +\frac{3}{2} \frac{1}{(z-w)^2} \, c \partial^2 c (w)
  +\frac{2}{3} \frac{1}{z-w} \, c \partial^3 c (w)
\nonumber \\ && {}
  -\frac{3}{2} \frac{1}{z-w} \, \partial c \partial^2 c (w)
  +\frac{1}{z-w} \, c \partial c T^m (w),
\end{eqnarray}
the BRST transformations of $c T^m (w)$
and $\no{ bc \partial c } (w)$ are given by
\begin{eqnarray}
  Q_B \cdot c T^m (w)
  &=& -\frac{13}{6} \, c \partial^3 c (w)
    - c \partial c T^m (w),
\nonumber \\
  Q_B \cdot \no{ bc \partial c } (w)
  &=& \frac{2}{3} \, c \partial^3 c (w)
    -\frac{3}{2} \, \partial c \partial^2 c (w)
    + c \partial c T^m (w).
\end{eqnarray}
As a check, we can verify that $Q_B \cdot j_B (w)$ vanishes from
these results.
The quantity $\langle \phi | Q_B | \Psi^{(2)} \rangle$
is then given by
\begin{eqnarray}
  \langle \phi | Q_B | \Psi^{(2)} \rangle
  &=& \frac{x}{\sqrt{1-t}} \vev{ \phi | B_t(c \partial c)}
  + \sqrt{1-t} \left[
    \left( u -\frac{13}{6} \, v + \frac{2}{3} \, w \right)
    \langle \phi | B_t(c \partial^3 c) \rangle \right.
\nonumber \\ && {} \left.
  + \left( u -\frac{3}{2} \, w \right)
    \langle \phi | B_t(\partial c \partial^2 c) \rangle
  + ( -v+w ) \vev{ \phi | B_t(c \partial c T^m) } \right]
\nonumber \\
  &=& \frac{x}{\sqrt{1-t}} \vev{ \phi | B(c \partial c)}
  + \sqrt{1-t} \left[
    \left( \frac{2}{3} \, x + u -\frac{13}{6} \, v
    + \frac{2}{3} \, w \right)
    \langle \phi | B(c \partial^3 c) \rangle \right.
\nonumber \\ && {} \left.
  + \left( -\frac{3}{2} \, x + u -\frac{3}{2} \, w \right)
    \langle \phi | B(\partial c \partial^2 c) \rangle
  + ( x-v+w ) \vev{ \phi | B(c \partial c T^m) } \right]
\nonumber \\ && {}
  + O( (1-t)^{\frac{3}{2}} ).
\label{phi-Q_B-Psi-2}
\end{eqnarray}

The computation of
$\langle \phi | \Psi^{(2)} \ast \Psi^{(2)} \rangle$
takes the same steps as in the case of the leading order.
First, we have to map the operators inserted at the origin
in the coordinate $z'= I \circ f_t (\xi)$
to $\pm 1/\beta$ in the $z$ coordinate
using the relation (\ref{z-z'}).
The operator $\partial^2 c$ is not a primary field,
but its conformal transformation law can be easily derived from
that of $c$. We will also need the transformation law
of $\partial c$ later. They are given by
\begin{eqnarray}
  \partial c ( f(z) ) &\to&
  \partial c (z)
  + \frac{f''(z)}{f'(z)} \, c (z),
\\
  \partial^2 c ( f(z) ) &\to&
  \frac{1}{f'(z)} \, \partial^2 c (z)
  + \frac{f''(z)}{f'(z)^2} \, \partial c (z)
  + \left\{ \frac{f'''(z)}{f'(z)^2}
  - \frac{f''(z)^2}{f'(z)^3} \right\} c(z).
\end{eqnarray}
The conformal transformation of $c T^m$ can be obtained
by evaluating the Schwarzian derivative
with $c=26$:
\begin{equation}
  c T^m ( f(z) ) \to
  \frac{1}{f'(z)} \, c T^m (z)
  + \left\{ - \frac{13}{6} \frac{f'''(z)}{f'(z)^2}
  + \frac{13}{4} \frac{f''(z)^2}{f'(z)^3} \right\} c(z).
\end{equation}
The last operator $\no{ bc \partial c}$ can be written as
a linear combination of $\partial^2 c$, $c T^m$,
and $j_B$. We have obtained the conformal transformations
of $\partial^2 c$ and $c T^m$, and the BRST current $j_B$
is a primary field:
\begin{equation}
  j_B ( f(z) ) \to \frac{1}{f'(z)} \, j_B (z).
\end{equation}
Therefore, the transformation law of $\no{ bc \partial c}$
is given by
\begin{equation}
  \no{ bc \partial c} ( f(z) ) \to
  \frac{1}{f'(z)} \, \no{ bc \partial c} (z)
  - \frac{3}{2} \frac{f''(z)}{f'(z)^2} \, \partial c (z)
  + \left\{ \frac{2}{3} \frac{f'''(z)}{f'(z)^2}
  - \frac{7}{4} \frac{f''(z)^2}{f'(z)^3} \right\} c(z).
\end{equation}
The derivatives $d^2 z'/dz^2$ and $d^3 z'/dz^3$
at $z= \pm 1/\beta$ can be computed from (\ref{z-z'}) as follows:
\begin{eqnarray}
  \left. \frac{d^2 z'}{dz^2} \right|_{z= \pm \frac{1}{\beta}}
  &=& \pm
  \frac{\beta^2 \, d^2 \, \left( 3 \, \beta^2 - 7 \, d^2 \right) \,
       {\sqrt{1 - t^4}}}
       {2 \, {\sqrt{3}} \, {\left( \beta^2 + d^2 \right) }^2},
\\
  \left. \frac{d^3 z'}{dz^3} \right|_{z= \pm \frac{1}{\beta}}
  &=& \frac{\beta^3 \, d^4 \,
      \left( 122 \, d^2 + 27 \, d^2 \, t^2 -120 \, \beta^2 \right) \,
      {\sqrt{1 - t^4}}}
      {8 \, {\sqrt{3}} \, {\left( \beta^2 + d^2 \right) }^3}.
\end{eqnarray}
The conformal transformations of $\partial^2 c$, $c T^m$,
and $\no{ bc \partial c }$ can be calculated by plugging these
derivatives into their transformation laws.

In Subsection 2.2, we made the conformal map $z/(z+p)$
to bring the infinity to a finite point in studying
the limit $t \to 1$. This is convenient when we explicitly compute
the inner product
$\langle \phi | \Psi^{(2)} \ast \Psi^{(2)} \rangle$
for a given $\ket{\phi}$ because the operator $\phi(0)$
in the original coordinate is also mapped to a finite point.
In solving the equation of motion, however,
it is sufficient to make a simple inversion
which brings the operators at $z=\pm 1/\beta$ to $\mp \beta$.
After the inversion,
the operators at $z'=0$ are mapped as follows:
\begin{eqnarray}
  && \frac{x}{\sqrt{1-t}} \, c(0)
  + \sqrt{1-t} \left[ u \, \partial^2 c (0)
  + v \, c T^m (0)
  + w \, \no{ bc \partial c } (0) \right]
\nonumber\\ &&
  \to \frac{1}{\beta} \left(
  \frac{2}{\sqrt{3}} \, x + \frac{55}{27 \sqrt{3}} \, u
  -\frac{247}{72 \sqrt{3}} \, v +\frac{\sqrt{3}}{8} \, w
  + O(1-t) \right) c (\mp \beta)
\nonumber \\ && \qquad
  {} \mp \left( -\frac{7}{6 \sqrt{3}} \, u
  +\frac{7}{4 \sqrt{3}} \, w + O(1-t) \right)
  \partial c (\mp \beta)
\nonumber \\ && \qquad
  {} + \beta \left( \frac{\sqrt{3}}{2} + O(1-t) \right)
  \left[ u \, \partial^2 c (\mp \beta)
  + v \, c T^m (\mp \beta)
  + w \, \no{ bc \partial c } (\mp \beta) \right].
\label{0-beta}
\end{eqnarray}
We have only presented the leading terms
in the limit $t \to 1$ here, but the expression for any $t$
is also available.
  Since $\beta$ goes to zero as $t \to 1$,
the operators at $\mp \beta$ can be expanded
in terms of operators at the origin.
If we define $V_{\pm} (z)$ by
\begin{equation}
{\renewcommand\arraystretch{1.2}
  V_{\pm} (z) \equiv
  \left( \begin{array}{c}
  \beta^{-1} c (z) \\
  \pm \partial c (z) \\
  \beta \, \partial^2 c (z) \\
  \beta \, c T^m (z) \\
  \beta \, \no{ bc \partial c } (z)
  \end{array} \right),
}
\end{equation}
relevant OPE's can be expressed and evaluated as follows:
\begin{eqnarray}
  && \frac{1}{2} \, V_{-} (-\beta) \, V_{+}^{T} (\beta)
     -\frac{1}{2} \, V_{+} (\beta) \, V_{-}^{T} (-\beta)
\nonumber \\
  && = \frac{1}{\beta} \, {\cal M}_1 \, c \partial c (0)
  + \beta \, {\cal M}_2 \, c \partial^3 c (0)
  + \beta \, {\cal M}_3 \, \partial c \partial^2 c (0)
  + \beta \, {\cal M}_4 \, c \partial c T^m (0)
\nonumber \\
  && \quad {} + O(\beta^3),
\label{beta-OPE}
\end{eqnarray}
where $V_{\pm}^{T} (z)$ denotes the transpose
of $V_{\pm} (z)$ and
\begin{eqnarray}
{\renewcommand\arraystretch{1.2}
  {\cal M}_1 = \left( \begin{array}{ccccc}
  2 & 1 & 0 & 0 & -\frac{1}{2} \\
  1 & 0 & 0 & 0 &  \frac{1}{4} \\
  0 & 0 & 0 & 0 & -\frac{1}{4} \\
  0 & 0 & 0 & \frac{13}{8} & 0 \\
  -\frac{1}{2} & \frac{1}{4} & -\frac{1}{4} & 0 & -\frac{3}{8}
  \end{array} \right), \quad
  {\cal M}_2 = \left( \begin{array}{ccccc}
  \frac{1}{3} & \frac{1}{2} & 1 & 0 & -\frac{1}{4} \\
  \frac{1}{2} & 0 & 0 & 0 &  \frac{1}{8} \\
  1 & 0 & 0 & 0 & -\frac{1}{8} \\
  0 & 0 & 0 & \frac{13}{48} & 0 \\
  -\frac{1}{4} & \frac{1}{8} & -\frac{1}{8} & 0 & -\frac{7}{48}
  \end{array} \right),
}
\nonumber \\
{\renewcommand\arraystretch{1.2}
  {\cal M}_3 = \left( \begin{array}{ccccc}
  -1 & -\frac{3}{2} & -1 & 0 & -\frac{1}{4} \\
  -\frac{3}{2} & -2 & -1 & 0 &  \frac{1}{8} \\
  -1 & -1 & 0 & 0 & -\frac{1}{8} \\
  0 & 0 & 0 & -\frac{13}{16} & 0 \\
  -\frac{1}{4} & \frac{1}{8} & -\frac{1}{8} & 0 & \frac{15}{16}
  \end{array} \right), \quad
  {\cal M}_4 = \left( \begin{array}{ccccc}
  0 & 0 & 0 & 2 & 0 \\
  0 & 0 & 0 & 1 & 0 \\
  0 & 0 & 0 & 0 & 0 \\
  2 & 1 & 0 & 1& -\frac{1}{2} \\
  0 & 0 & 0 & -\frac{1}{2} & 0
  \end{array} \right).
}
\end{eqnarray}
Therefore, the quantity
$\langle \phi | \Psi^{(2)} \ast \Psi^{(2)} \rangle$
can be expanded as follows:
\begin{eqnarray}
  \langle \phi | \Psi^{(2)} \ast \Psi^{(2)} \rangle
  &=& \frac{{\cal F}_1}{\sqrt{1-t}}
    \langle{ \phi | \widetilde{B}_t (c \partial c) \rangle}
    + \sqrt{1-t} \left[
    {\cal F}_2 \langle \phi
    | \widetilde{B}_t (c \partial^3 c) \rangle \right.
    + {\cal F}_3 \langle \phi
    | \widetilde{B}_t (\partial c \partial^2 c) \rangle
\nonumber \\ && \left. {}
    + {\cal F}_4 \langle \phi
    | \widetilde{B}_t (c \partial c T^m) \rangle \right]
    + O \left( (1-t)^{\frac{3}{2}} \right),
\end{eqnarray}
where ${\cal F}_1$, ${\cal F}_2$, ${\cal F}_3$, and ${\cal F}_4$
are functions of $x$, $u$, $v$, and $w$, and we have defined
\begin{equation}
  \langle \phi | \widetilde{B}_t ({\cal O}) \rangle
  = \vev{ I \circ \widetilde{f}_t \circ \phi (0) \, {\cal O}(0) }
\end{equation}
for any state $\ket{\phi}$ in the Fock space. From
(\ref{0-beta}) and (\ref{beta-OPE}),
the functions ${\cal F}_1$, ${\cal F}_2$, ${\cal F}_3$,
and ${\cal F}_4$ can be computed.
For example, ${\cal F}_1$ is given by
\begin{eqnarray}
  {\cal F}_1 = \frac{3^{\frac{3}{4}}}{\sqrt{2}} \left(
  \frac{8\,x^2}{3}
  + \frac{314\,x\,u}{81}
  + \frac{2585\,u^2}{2187}
  - \frac{247\,x\,v}{27}
  - \frac{38779\,u\,v}{5832}
  + \frac{35243\,v^2}{3888} \right.
\nonumber \\ \left. {}
  + \frac{7\,x\,w}{3}
  + \frac{919\,u\,w}{648}
  - \frac{1729\,v\,w}{432}
  + \frac{w^2}{2} \right) + O(1-t).
\end{eqnarray}
The leading part of the equation
$\langle \phi | Q_B | \Psi^{(2)} \rangle
+ \langle \phi | \Psi^{(2)} \ast \Psi^{(2)} \rangle = 0$,
which is proportional to $\vev{\phi | B(c \partial c)}$
and is of $O(1/\sqrt{1-t})$, is therefore given by
\begin{eqnarray}
  x + \frac{3^{\frac{3}{4}}}{\sqrt{2}} \left(
  \frac{8\,x^2}{3}
  + \frac{314\,x\,u}{81}
  + \frac{2585\,u^2}{2187}
  - \frac{247\,x\,v}{27}
  - \frac{38779\,u\,v}{5832}
  + \frac{35243\,v^2}{3888} \right.
\nonumber \\ \left. {}
  + \frac{7\,x\,w}{3}
  + \frac{919\,u\,w}{648}
  - \frac{1729\,v\,w}{432}
  + \frac{w^2}{2} \right) = 0.
\label{equation-1}
\end{eqnarray}
As we mentioned, there are two important sources
of terms at the next-to-leading order.
The first kind of terms come from ${\cal F}_2$, 
${\cal F}_3$, and ${\cal F}_4$.
The second kind of terms which come from
the difference between $f_t (\xi)$ and $\widetilde{f}_t (\xi)$
at $O(1-t)$ can be taken into account
by replacing $x$ with $(1-1/\sqrt{3}) \, x$
in the coefficients in front of
$\langle \phi | B(c \partial^3 c) \rangle$,
$\langle \phi | B(\partial c \partial^2 c) \rangle$,
and $\vev{ \phi | B(c \partial c T^m) }$
in (\ref{phi-Q_B-Psi-2})
because of the equation (\ref{equation-1}).
By computing ${\cal F}_2$, ${\cal F}_3$, and ${\cal F}_4$ from
(\ref{0-beta}) and (\ref{beta-OPE}),
we obtain the following three equations
at the next-to-leading order:
\begin{eqnarray}
  && \frac{2}{3} \left( 1 - \frac{1}{\sqrt{3}} \right) x
  + u -\frac{13 \, v}{6} + \frac{2 \, w}{3}
  + \frac{\sqrt{2}}{3^{\frac{3}{4}}} \left(
  \frac{4\,x^2}{9}
  + \frac{517\,x\,u}{243}
  + \frac{22385\,u^2}{13122}
  - \frac{247\,x\,v}{162} \right.
\nonumber \\ && \quad \left. {}
  - \frac{127699\,u\,v}{34992}
  + \frac{35243\,v^2}{23328}
  + \frac{5\,x\,w}{6}
  + \frac{965\,u\,w}{1296}
  - \frac{1235\,v\,w}{864}
  + \frac{w^2}{4} \right) = 0,
\label{equation-2} \\
  && -\frac{3}{2} \left( 1 - \frac{1}{\sqrt{3}} \right) x
  + u -\frac{3 \, w}{2}
  + \frac{\sqrt{2}}{3^{\frac{3}{4}}} \left(
  - \frac{4\,x^2}{3}
  - \frac{193\,x\,u}{81}
  - \frac{3431\,u^2}{4374}
  + \frac{247\,x\,v}{54} \right.
\nonumber \\ && \quad \left. {}
  + \frac{47671\,u\,v}{11664}
  - \frac{35243\,v^2}{7776}
  - \frac{9\,x\,w}{2}
  - \frac{559\,u\,w}{144}
  + \frac{247\,v\,w}{32}
  - \frac{23\,w^2}{12} \right) = 0,
\label{equation-3} \\
  && \left( 1 - \frac{1}{\sqrt{3}} \right) x -v +w
  + \frac{\sqrt{2}}{3^{\frac{3}{4}}} \left(
  4\,x\,v
  + \frac{157\,u\,v}{54}
  - \frac{55\,v^2}{9}
  + \frac{7\,v\,w}{4} \right) = 0.
\label{equation-4}
\end{eqnarray}
There are also terms of $O(\sqrt{1-t})$
which are proportional to $\vev{ \phi | B(c \partial c) }$.
The equation coming from $\vev{ \phi | B(c \partial c) }$
at this subleading order can be easily satisfied
by introducing a subleading part of $x$.
We will not compute it because, as we will see,
it does not contribute to the energy
density of the solution in the limit $t \to 1$.

We have the four equations (\ref{equation-1}), (\ref{equation-2}),
(\ref{equation-3}), and (\ref{equation-4}) to solve
for the four variables $x$, $u$, $v$, and $w$.
We could not solve them analytically,
but we found six nontrivial, real-valued solutions
numerically by using {\it Mathematica}.
They are given in Table \ref{solutions}.
\begin{table}[ht]
\caption{Solutions at the next-to-leading order.}
\label{solutions}
\begin{center}
{\renewcommand\arraystretch{1.1}
\begin{tabular}{|c|c|c|c|}
  \hline
  $x$ & $u$ & $v$ & $w$ \\
  \hline
  $-3.94824$ & $28.5424$ & $15.5319$ & $15.9191$ \\
  \hline
  $8.23295$ & $-10.1642$ & $-0.83021$ & $0.450412$ \\
  \hline
  $-0.170713$ & $1.19342$ & $0.726956$ & $0.863933$ \\
  \hline
  $-0.241241$ & $-0.486967$ & $-0.341072$ & $-0.479342$ \\
  \hline
  $-0.329909$ & $0.00107446$ & $0.00924674$ & $0.155003$ \\
  \hline
  $-0.0000505996$ & $0.0138804$ & $0.00924674$ & $0.00926889$ \\
  \hline
\end{tabular}
}
\end{center}
\end{table}
The numerical value of $x$ at the leading-order solution is
\begin{equation}
  x_{{\rm leading}}
  = - \frac{3^{\frac{1}{4}}}{4 \sqrt{2}} \simeq -0.232651.
\end{equation}
The value of $x$ in the fourth solution is
fairly close to this value,
and those of the third and fifth ones are not too far away.
At this point, however, there are no proper reasons
to select a subset among the six.
In the next subsection, we will evaluate the energy density
for these solutions, and it turns out that
the fourth and fifth solutions are selected from
the condition (ii-a) in the Introduction.

\subsection{Energy density at the next-to-leading order}

As we did for the leading-order solution $| \Psi^{(0)} \rangle$
in Subsection 2.3, let us compute
$\langle \Psi^{(2)} | Q_B | \Psi^{(2)} \rangle$
and $\langle \Psi^{(2)} | \Psi^{(2)} \ast \Psi^{(2)} \rangle$.
It will turn out to be convenient
to introduce the following parametrization
of $| \Psi^{(2)} \rangle$ for the computations of these quantities:
\begin{eqnarray}
  | \Psi^{(2)} \rangle
  &=& \frac{\tilde{x}}{\sqrt{1-t^4}} \ket{B_t(c)}
\nonumber \\
  && {}
  + \sqrt{1-t^4} \left[ \tilde{u} \, | B_t(\partial^2 c) \rangle 
  + \tilde{v} \ket{ B_t(c T^m) }
  + \tilde{w} \ket{ B_t(\no{ bc \partial c}) } \right].
\end{eqnarray}

Let us start with
$\langle \Psi^{(2)} | Q_B | \Psi^{(2)} \rangle$. From
the relation (\ref{z-z'-2}), the derivatives
$d^2 z'/dz^2$ and $d^3 z'/dz^3$ at $z = \pm 1$
can be computed as follows:
\begin{equation}
  \left. \frac{d^2 z'}{dz^2} \right|_{z = \pm 1}
  = \mp \frac{\sqrt{1-t^4}}{2}, \quad
  \left. \frac{d^3 z'}{dz^3} \right|_{z = \pm 1}
  = \frac{3 \, (2 + t^2) \sqrt{1-t^4}}{8}.
\end{equation}
The operators inserted at $z'=0$ are mapped
to $z=-1$ in the $z$ coordinate as
\begin{eqnarray}
  && \frac{\tilde{x}}{\sqrt{1-t^4}} \, c (0)
  + \sqrt{1-t^4} \left\{ \tilde{u} \, \partial^2 c (0)
  + \tilde{v} \, c T^m (0)
  + \tilde{w} \, \no{ bc \partial c} (0) \right\}
\nonumber \\
  &\to& \left( \frac{\tilde{x}}{2}
  + \frac{2 + 3 \, t^2}{2} \tilde{u}
  - \frac{13 \, t^2}{4} \tilde{v}
  + \frac{2 \, t^2 - 3}{2} \tilde{w} \right) c (-1)
  + ( 2 \, \tilde{u} - 3 \, \tilde{w} ) \, \partial c (-1)
\nonumber \\ && {}
  + 2 \, \tilde{u} \, \partial^2 c (-1)
  + 2 \, \tilde{v} \, c T^m (-1)
  + 2 \, \tilde{w} \, \no{ bc \partial c} (-1).
\end{eqnarray}
The BRST transformations of the operators at $z'=0$
are mapped to $z=1$ in the $z$ coordinate as
\begin{eqnarray}
  && Q_B \cdot \left[ \frac{\tilde{x}}{\sqrt{1-t^4}} \, c (0)
  + \sqrt{1-t^4} \left\{ \tilde{u} \, \partial^2 c (0)
  + \tilde{v} \, c T^m (0)
  + \tilde{w} \, \no{ bc \partial c} (0) \right\} \right]
\nonumber \\
  &=& \left( \frac{\tilde{x}}{2}
  + \frac{2 + 3 \, t^2}{2} \tilde{u}
  - \frac{13 \, t^2}{4} \tilde{v}
  + \frac{2 \, t^2 - 3}{2} \tilde{w} \right) c \partial c (1)
  + ( -2 \, \tilde{u} + 3 \, \tilde{w} ) \, c \partial^2 c (1)
\nonumber \\ && {}
  + \left( 2 \, \tilde{u}
  - \frac{13}{3} \, \tilde{v}
  + \frac{4}{3} \, \tilde{w} \right) c \partial^3 c (1)
  + ( 2 \, \tilde{u} - 3 \, \tilde{w} ) \,
    \partial c \partial^2 c (1)
  + ( -2 \, \tilde{v} + 2 \, \tilde{w} ) \, c \partial c T^m (1).
\nonumber \\
\end{eqnarray}
It is tedious but straightforward to compute
the correlation functions of these operators.
The final result for
$\langle \Psi^{(2)} | Q_B | \Psi^{(2)} \rangle$
is given by
\begin{eqnarray}
  \langle \Psi^{(2)} | Q_B | \Psi^{(2)} \rangle_{density}
  &=& -\tilde{x}^2
  - 6 \, t^2 \, \tilde{x} \, \tilde{u}
  - ( 4 + 9 \, t^4 ) \, \tilde{u}^2
  + 13 \, t^2 \, \tilde{x} \,\tilde{v}
  + 39 \, t^4 \, \tilde{u} \, \tilde{v}
\nonumber \\ && {}
  + \left( 13 - \frac{169 \, t^4}{4} \right) \tilde{v}^2
  - 4 \, t^2 \, \tilde{x} \, \tilde{w}
  + 12 \, (1-t^4) \, \tilde{u} \, \tilde{w}
\nonumber \\ && {}
  - 26 \, (1-t^4) \, \tilde{v} \, \tilde{w}
  + 4 \, (1-t^4) \, \tilde{w}^2.
\end{eqnarray}
We are interested in the limit $t \to 1$.
As can be seen from this expression, the quantity
$\langle \Psi^{(2)} | Q_B | \Psi^{(2)} \rangle$
has a finite limit as $t$ goes to $1$.
Since
\begin{equation}
  \lim_{t \to 1} \tilde{x} = 2 \, x, \quad
  \lim_{t \to 1} \tilde{u} = \frac{1}{2} \, u, \quad
  \lim_{t \to 1} \tilde{v} = \frac{1}{2} \, v, \quad
  \lim_{t \to 1} \tilde{w} = \frac{1}{2} \, w,
\end{equation}
it is given as follows:
\begin{eqnarray}
  && \lim_{t \to 1}
  \langle \Psi^{(2)} | Q_B | \Psi^{(2)} \rangle_{density}
\nonumber \\
  && = -4\,x^2 - 6\,x\,u - \frac{13\,u^2}{4} + 13\,x\,v + 
  \frac{39\,u\,v}{4} - \frac{117\,v^2}{16} - 4\,x\,w.
\end{eqnarray}

The computation of
$\langle \Psi^{(2)} | \Psi^{(2)} \ast \Psi^{(2)} \rangle$
can be done in a similar way. From the relation (\ref{z-z'-3}),
the derivatives $d^2 z'/dz^2$ and $d^3 z'/dz^3$
at $z = 0$ and $z = \pm \sqrt{3}$
can be computed as follows:
\begin{eqnarray}
  && \left. \frac{d^2 z'}{dz^2} \right|_{z = 0} = 0, \quad
  \left. \frac{d^3 z'}{dz^3} \right|_{z = 0}
  = \frac{3 \, (10 + 27 \, t^2) \sqrt{1-t^4}}{8},
\nonumber \\
 && \left. \frac{d^2 z'}{dz^2} \right|_{z = \pm \sqrt{3}}
  = \mp \frac{3 \sqrt{3 \, (1-t^4)}}{16}, \quad
  \left. \frac{d^3 z'}{dz^3} \right|_{z = \pm \sqrt{3}}
  = \frac{3 \, (82 + 27 \, t^2) \sqrt{1-t^4}}{512}.
\end{eqnarray}
The operators inserted at $z'=0$ are mapped
to $z=0$ and $z = \pm \sqrt{3}$ in the $z$ coordinate as
\begin{eqnarray}
  && \frac{\tilde{x}}{\sqrt{1-t^4}} \, c (0)
  + \sqrt{1-t^4} \left\{ \tilde{u} \, \partial^2 c (0)
  + \tilde{v} \, c T^m (0)
  + \tilde{w} \, \no{ bc \partial c} (0) \right\}
\nonumber \\
  &\to& \left( \frac{3}{2} \, \tilde{x}
  + \frac{10 + 27 \, t^2}{6} \tilde{u}
  - \frac{13 \, (10 + 27 \, t^2)}{36} \tilde{v}
  + \frac{10 + 27 \, t^2}{9} \tilde{w} \right) c (0)
\nonumber \\ && {}
  + \frac{2}{3} \, \tilde{u} \, \partial^2 c (0)
  + \frac{2}{3} \, \tilde{v} \, c T^m (0)
  + \frac{2}{3} \, \tilde{w} \, \no{ bc \partial c} (0),
\nonumber \\
  && \left( \frac{3}{8} \, \tilde{x}
  + \frac{34 + 27 \, t^2}{24} \tilde{u}
  - \frac{13 \, (10 + 27 \, t^2)}{144} \tilde{v}
  - \frac{44 - 27 \, t^2}{36} \tilde{w} \right) c (\pm \sqrt{3})
\nonumber \\ && {}
  \pm \left( -\frac{4 \sqrt{3}}{3} \, \tilde{u}
  + 2 \sqrt{3} \, \tilde{w} \right) \partial c (\pm \sqrt{3})
\nonumber \\ && {}
  + \frac{8}{3} \, \tilde{u} \, \partial^2 c (\pm \sqrt{3})
  + \frac{8}{3} \, \tilde{v} \, c T^m (\pm \sqrt{3})
  + \frac{8}{3} \, \tilde{w} \, \no{ bc \partial c} (\pm \sqrt{3}).
\end{eqnarray}
It is again tedious but straightforward to compute
the correlation functions of these operators.
The final expression of
$\langle \Psi^{(2)} | \Psi^{(2)} \ast \Psi^{(2)} \rangle$
for an arbitrary $t$ is given in Appendix B.
As in the case of
$\langle \Psi^{(2)} | Q_B | \Psi^{(2)} \rangle$,
it also has a finite limit as $t$ goes to $1$, which is given by
\begin{eqnarray}
  && \lim_{t \to 1}
  \langle \Psi^{(2)} | \Psi^{(2)} \ast \Psi^{(2)} \rangle_{density}
  = -\frac{81\,{\sqrt{3}}\,x^3}{8}
  -\frac{927\,{\sqrt{3}}\,x^2\,u}{32}
  -\frac{3451\,{\sqrt{3}}\,x\,u^2}{128}
\nonumber \\ && \quad {}
  -\frac{4205\,{\sqrt{3}}\,u^3}{512}
  +\frac{4329\,{\sqrt{3}}\,x^2\,v}{64}
  +\frac{49543\,x\,u\,v}{128\,{\sqrt{3}}}
  +\frac{1659931\,u^2\,v}{9216\,{\sqrt{3}}}
  -\frac{244673\,x\,v^2}{512\,{\sqrt{3}}}
\nonumber \\ && \quad {}
  -\frac{25201319\,u\,v^2}{55296\,{\sqrt{3}}}
  +\frac{43213963\,v^3}{110592\,{\sqrt{3}}}
  -\frac{315\,{\sqrt{3}}\,x^2\,w}{16}
  -\frac{1095\,{\sqrt{3}}\,x\,u\,w}{32}
\nonumber \\ && \quad {}
  -\frac{286195\,u^2\,w}{6912\,{\sqrt{3}}}
  +\frac{16835\,x\,v\,w}{64\,{\sqrt{3}}}
  +\frac{175565\,u\,v\,w}{768\,{\sqrt{3}}}
  -\frac{8563555\,v^2\,w}{27648\,{\sqrt{3}}}
\nonumber \\ && \quad {}
  -\frac{403\,{\sqrt{3}}\,x\,w^2}{32}
  -\frac{104687\,u\,w^2}{3456\,{\sqrt{3}}}
  +\frac{193843\,v\,w^2}{2304\,{\sqrt{3}}}
  -\frac{169\,{\sqrt{3}}\,w^3}{64}.
\end{eqnarray}

We can now evaluate ${\cal R} [ \Psi^{(2)} ]$
and ${\cal E} [ \Psi^{(2)} ]/T_{25}$ in the limit $t \to 1$
for the numerical solutions
we found in the previous subsection.
The result is summarized in Table \ref{ratio-tension}.
\begin{table}[ht]
\caption{${\cal R}$ and ${\cal E}/T_{25}$
in the limit $t \to 1$.}
\label{ratio-tension}
\begin{center}
{\renewcommand\arraystretch{1.1}
\begin{tabular}{|c||c||c|c|c|c|}
  \hline
  ${\cal R}$ & ${\cal E}/T_{25}$ & $x$ & $u$ & $v$ & $w$ \\
  \hline
  $0.143959$ & $-1188.83$
  & $-3.95$ & $28.5$ & $15.5$ & $15.9$ \\
  \hline
  $-0.61934$ & $98.9891$
  & $8.23$ & $-10.2$ & $-0.830$ & $0.450$ \\
  \hline
  $-0.822535$ & $0.0897315$
  & $-0.171$ & $1.19$ & $0.727$ & $0.864$ \\
  \hline
  $-0.998109$ & $-1.08981$
  & $-0.241$ & $-0.487$ & $-0.341$ & $-0.479$ \\
  \hline
  $-0.998909$ & $-0.882631$
  & $-0.330$ & $0.00107$ & $0.00925$ & $0.155$ \\
  \hline
  $-0.326083$ & $3.05927 \times {10}^{-8}$ 
  & $-0.0000506$ & $0.0139$ & $0.00925$ & $0.00927$ \\
  \hline
\end{tabular}
}
\end{center}
\end{table}
The values of ${\cal R}$ for
the fourth and fifth solutions are very close to $-1$.
Therefore, these solutions approximately satisfy the equation
of motion even when it is contracted with the solution itself.
Furthermore, the values of ${\cal E}/T_{25}$
for these two solutions are fairly close to $-1$.
They give approximately 109\% and 88\%, respectively,
of the D25-brane tension.
Compared with the solution at the leading order
which gives approximately 68\% of the D25-brane tension,
these values of the energy density for these solutions
at the next-to-leading order are closer to
the predicted value $- T_{25}$.

It is not surprising that we have found more than one
numerical solution at the next-to-leading order
because we do not fix gauge.
Multiple numerical solutions were also found
in the level truncation analysis without gauge fixing
by Ellwood and Taylor \cite{Ellwood:2001ne},
and there are two solutions at level 2.
In fact, the values of the energy density
for our solutions are interestingly close to those
for the two solutions at level 2 in level truncation.
We will review the results of level truncation
in the next subsection, and compare them with ours.

\subsection{Comparison with solutions by a variational method}

A state in the form of $\ket{B_t(\cal O)}$
reduces to a state $\ket{\cal O}$ in the Fock space
in the limit $t \to 0$.
Therefore, the results of the ordinary level-truncation analysis
can be reproduced simply by taking the limit $t \to 0$ of ours.
In this limit, the parameters $\tilde{x}$, $\tilde{u}$,
$\tilde{v}$, and $\tilde{w}$ reduce to $x$, $u$, $v$, and $w$,
respectively, so that ${\cal E}/T_{25}$ in the limit is
given by
\begin{eqnarray}
  && \left. \frac{{\cal E}}{T_{25}} \right|_{t=0}
  = 2 \pi^2 \left(
  - \frac{x^2}{2} - 2\,u^2 + \frac{13\,v^2}{2}
  + 6\,u\,w - 13\,v\,w + 2\,w^2 \right.
\nonumber \\ && \quad {}
  -\frac{27\,{\sqrt{3}}\,x^3}{64} - 
  \frac{33\,{\sqrt{3}}\,x^2\,u}{32} - 
  \frac{19\,x\,u^2}{16\,{\sqrt{3}}} - 
  \frac{u^3}{8\,{\sqrt{3}}} + 
  \frac{195\,{\sqrt{3}}\,x^2\,v}{64} + 
  \frac{715\,x\,u\,v}{48\,{\sqrt{3}}}
\nonumber \\ && \quad {}
  + \frac{1235\,u^2\,v}{432\,{\sqrt{3}}} - 
  \frac{7553\,x\,v^2}{192\,{\sqrt{3}}} - 
  \frac{83083\,u\,v^2}{2592\,{\sqrt{3}}} + 
  \frac{272363\,v^3}{5184\,{\sqrt{3}}} - 
  \frac{3\,{\sqrt{3}}\,x^2\,w}{4} + {\sqrt{3}}\,x\,u\,w
\nonumber \\ && \quad {} \left.
  + \frac{703\,u^2\,w}{81\,{\sqrt{3}}} + 
  \frac{65\,x\,v\,w}{6\,{\sqrt{3}}} - 
  \frac{65\,u\,v\,w}{9\,{\sqrt{3}}} - 
  \frac{7553\,v^2\,w}{324\,{\sqrt{3}}} - 
  \frac{x\,w^2}{{\sqrt{3}}} + 
  \frac{94\,u\,w^2}{81\,{\sqrt{3}}} + 
  \frac{65\,v\,w^2}{27\,{\sqrt{3}}} \right).
\nonumber \\
\end{eqnarray}
In the level-truncation analysis
by Sen and Zwiebach \cite{Sen:1999nx},
the Siegel gauge condition was imposed.
It corresponds to setting the parameter $w$ to zero
in our expression.
The expression in \cite{Sen:1999nx} at level 2
is precisely reproduced
by the following identification of the parameters:
\begin{equation}
  x_{ours} = - t_{SZ}, \quad
  u_{ours} = - \frac{1}{2} \, u_{SZ}, \quad
  v_{ours} = - \frac{1}{\sqrt{13}} \, v_{SZ}, \quad
  w_{ours} = 0.
\end{equation}
By taking variations with respect to $x$, $u$, and $v$,
we numerically find the following stationary point:
\begin{equation}
  x_c \simeq -0.544204, \quad
  u_c \simeq -0.0950952, \quad
  v_c \simeq -0.0559637,
\end{equation}
which reproduces the result by Sen and Zwiebach
at level 2 in the Siegel gauge:
\begin{equation}
 \frac{{\cal E}}{T_{25}} \simeq -0.959377.
\end{equation}

The level-truncation analysis can be done
without imposing a gauge condition,
and it was studied by Rastelli and Zwiebach
\cite{Rastelli:2000iu}
and by Ellwood and Taylor \cite{Ellwood:2001ne}.
By taking variations with respect to all of
$x$, $u$, $v$, and $w$,
we numerically find eight stationary points, which
we present in Table \ref{Ellwood-Taylor}
together with the values of ${\cal E}/T_{25}$.
The values of ${\cal E}/T_{25}$
for the fifth and sixth solutions are relatively close to
$-1$, giving approximately 108\% and 88\%
of the D25-brane tension, respectively.
They are the two solutions found by Ellwood and Taylor
in \cite{Ellwood:2001ne}.\footnote
{
One of them with 88\% of the D25-brane tension was first found
by Rastelli and Zwiebach \cite{Rastelli:2000iu}.
}
These values of the energy density are interestingly
close to those of our solutions at the next-to-leading order.
In the ordinary level-truncation analysis without gauge fixing,
there are no proper reasons to select the two solutions
among the eight.
In our case, we emphasize that we chose the two solutions
in Subsection 4.2
by the condition (ii-a), not by the criterion that
they give a better value for the energy density.

\begin{table}[ht]
\caption{Level truncation at level 2 without gauge fixing.}
\label{Ellwood-Taylor}
\begin{center}
{\renewcommand\arraystretch{1.1}
\begin{tabular}{|c||c|c|c|c|}
  \hline
  ${\cal E}/T_{25}$ & $x$ & $u$ & $v$ & $w$ \\
  \hline
  $-64.2501$ & $-1.84905$ & $9.8938$ & $5.12224$ & $5.24925$ \\
  \hline
  $-103.454$ & $1.83562$ & $-3.53656$ & $-1.04338$ & $-0.455299$ \\
  \hline
  $0.212599$ & $-0.212556$ & $0.732749$ & $0.440045$ & $0.542256$ \\
  \hline
  $1303.36$ & $0.551333$ & $-0.63015$ & $-1.4288$ & $-14.0212$ \\
  \hline
  $-1.0778$ & $-0.590367$ & $-0.283678$ & $-0.191173$ & $-0.268326$ \\
  \hline
  $-0.88015$ & $-0.572474$ & $0.053691$ & $0.0422479$ & $0.180675$ \\
  \hline
  $8.16109 \times {10}^{-11}$ &
  $0.$ & $1.19631 \times {10}^{-6}$ &
  $0$ & $7.97539 \times {10}^{-7}$ \\
  \hline
  $8.16104 \times {10}^{-11}$ &
  $0.$ & $-1.19631 \times {10}^{-6}$ &
  $0$ & $-7.97538 \times {10}^{-7}$ \\
  \hline
\end{tabular}
}
\end{center}
\end{table}

We can also perform a similar variational analysis
based on the expression for the energy density
in the limit $t \to 1$.
We numerically found six stationary points
listed in Table \ref{another-level-truncation}
together with the values of ${\cal E}/T_{25}$
for those solutions.
As can be seen from the table,
the values of ${\cal E}/T_{25}$ are precisely the same
up to the order we computed
as those for the first six solutions
in Table \ref{Ellwood-Taylor}.
This seems to indicate that
the set of configurations in the limit $t \to 1$
can be obtained from that of $t=0$ by field redefinition.
Note, however, that our solutions at the next-to-leading order
in Subsection 4.2
cannot be obtained simply by field redefinition from
the solutions of level truncation at level 2,
which can be seen from the fact that
the values of the energy density are close,
but not exactly the same.

\begin{table}[ht]
\caption{Variational analysis based on the expression
in the limit $t \to 1$.}
\label{another-level-truncation}
\begin{center}
{\renewcommand\arraystretch{1.1}
\begin{tabular}{|c||c|c|c|c|}
  \hline
  ${\cal E}_c/T_{25}$ & $x$ & $u$ & $v$ & $w$ \\
  \hline
  $1303.36$ & $10.5984$ & $-1.2603$ & $-2.8576$ & $-28.0423$ \\
  \hline
  $-64.2501$ & $-4.36718$ & $19.7876$ & $10.2445$ & $10.4985$ \\
  \hline
  $-103.454$ & $3.28698$ & $-7.07313$ & $-2.08675$ & $-0.910597$ \\
  \hline
  $-0.88015$ & $-0.410143$ & $0.107382$ & $0.0844958$ & $0.361351$ \\
  \hline
  $0.212599$ & $-0.317509$ & $1.4655$ & $0.880091$ & $1.08451$ \\
  \hline
  $-1.0778$ & $-0.222653$ & $-0.567355$ & $-0.382346$ & $-0.536653$ \\
  \hline
\end{tabular}
}
\end{center}
\end{table}

\section{Discussion}
\setcounter{equation}{0}

We solved the equation of motion
of Witten's string field theory up to $O((1-t)^{3/2})$
based on our ansatz using the regulated butterfly state.
The leading-order solution $| \Psi^{(0)} \rangle$ gives
approximately 68\% of the D25-brane tension,
and the two numerical solutions
at the next-to-leading order
give approximately 88\% and 109\%
of the D25-brane tension, respectively.
These values for the energy density
are close to those obtained by level truncation up to level 2
without gauge fixing \cite{Rastelli:2000iu, Ellwood:2001ne}.
Since we have studied only the first two orders
of our approximation scheme,
it is premature to speculate whether or not
our solution will converge
to the exact solution as we increase the order
of the approximation, but
we regard the results we have obtained so far
as encouraging.

We found two solutions at the next-to-leading order,
and we expect that the number of solutions will increase
as we improve our ansatz by taking into account higher-order terms.
Since we are solving an equation of motion,
gauge fixing is not necessary.
But in level truncation, the convergence of
the energy density is generally better when we impose
a gauge-fixing condition \cite{Rastelli:2000iu, Ellwood:2001ne}.
In particular, it is known from experience that
the Siegel gauge condition works well.
It would be therefore interesting to incorporate
gauge fixing into our approach.
We should note, however, that the leading-order solution
$| \Psi^{(0)} \rangle$ does not satisfy the Siegel gauge condition.
Furthermore, all of the four terms of $| \Psi^{(2)} \rangle$
in (\ref{Psi^(2)}) are necessary to have a nontrivial solution
to the set of the four equations
(\ref{equation-1}), (\ref{equation-2}),
(\ref{equation-3}), and (\ref{equation-4}) so that
we do not have any obvious ways to incorporate gauge fixing
into our approach.

The idea to solve the equation of motion
of Witten's string field theory using a star-algebra projector
is not new, and in fact it was the original motivation
to construct a star-algebra projector in \cite{Kostelecky:2000hz}.
It is also natural in the half-string picture
\cite{Rastelli:2001rj, Gross:2001rk, Kawano:2001fn, Gross:2001yk,
Furuuchi:2001df} or in the Moyal star formulation
of string field theory
\cite{Bars:2001ag, Bars:2002bt, Douglas:2002jm, Bars:2002nu,
Bars:2002qt, Bars:2002yj, Bars:2003cr, Bars:2003gu}.
In a sense, our ansatz can be regarded as
a well-defined way to regularize the subtleties
at the open-string midpoint in the half-string picture.
The wave functional of
the class of string fields $\ket{B_t(\cal O)}$
approximately factorizes into the left and right halves
as $t$ approaches 1,
and the degree of freedom at the open-string midpoint
is taken into account by the operator insertion $\cal O$.
We can also consider more general
ways to insert operators into the regulated butterfly state.
For example, we can consider multiple operator insertions.
As was shown in \cite{Okawa:2003cm},
the twisted regulated butterfly state
can be represented in this way.
We can further smear out the operator insertions
along the boundary.
This kind of generalization would be
one possible direction to be explored in the future.

As we have seen in Section 3, our approach is
applicable not only to Witten's string field theory
but also to vacuum string field theory.
It may also be useful in constructing a solution
which has a well-defined expression for the energy density
in string field theory with a different class of
kinetic operators constructed in \cite{Takahashi:2002ez}
and studied in \cite{Kishimoto:2002xi, Drukker:2002ct,
Drukker:2003hh, Takahashi:2003pp, Takahashi:2003xe}.

\section*{Acknowledgments}
I would like to thank Wati Taylor and Barton Zwiebach
for useful discussions.
This work was supported in part by
the DOE grants DF-FC02-94ER40818 (MIT)
and DE-FG03-92ER40701 (Caltech),
and by a McCone Fellowship in Theoretical Physics from
California Institute of Technology.


\newpage
\appendix
\renewcommand{\thesection}{Appendix \Alph{section}.}
\renewcommand{\theequation}{\Alph{section}.\arabic{equation}}

\section{Conformal field theory formulation of string field theory}
\setcounter{equation}{0}

In the CFT formulation of string field theory
\cite{LeClair:1988sp, LeClair:1988sj},
an open string field is represented
as a wave functional obtained by a path integral
over a certain region in a Riemann surface.
For example, a state $\ket{\phi}$ in the Fock space
can be represented as a wave functional on the arc
$| \xi |=1$ in an upper-half complex plane of $\xi$
by path-integrating over the interior of
the upper half of the unit disk $| \xi | < 1$
with the corresponding operator $\phi (0)$
inserted at the origin
and with the boundary condition of the open string
imposed on the part of the real axis $-1 \le \xi \le 1$.
A more general class of states
such as the regulated butterfly state
can be defined by a path integral over a different region of
a Riemann surface with a boundary
and with possible operator insertions.
When we parametrize the open string on the arc
as $\xi = e^{i \theta}$ with $0 \le \theta \le \pi$,
we refer to the region $\pi/2 \le \theta \le \pi$
as the left half of the open string,
and to the region $0 \le \theta \le \pi/2$
as the right half of the open string.
We also refer to the point $\theta=\pi/2$
as the open-string midpoint.

We use the standard definitions \cite{Witten:1985cc} of
the inner product $\vev{\phi_1 | \phi_2}$
and the star product $\ket{\phi_1 \ast \phi_2}$.
The state $\ket{\phi_1 \ast \phi_2}$ is defined by
gluing together
the right half of the open string of $\ket{\phi_1}$
and the left half of the open string of $\ket{\phi_2}$.
Gluing can be made by conformal transformations
which map the two regions to be glued together
into the same region.
The inner product $\vev{\phi_1 | \phi_2}$ is defined
by gluing the left and right halves of
the open string of $\ket{\phi_1 \ast \phi_2}$.

We use the doubling trick throughout the paper.
For example, $bc$ ghosts on an upper-half plane
are extended to the lower-half plane by
$c (\bar{z}) = \tilde{c} (z)$ and $b (\bar{z}) = \tilde{b} (z)$.
The normalization of correlation functions is given by
\begin{equation}
  \vev{ c(z_1) \, c(z_2) \, c(z_3) }
  = ( z_1 - z_2 )( z_1 - z_3 )( z_2 - z_3 ) \int d^{26} x.
\label{three-c}
\end{equation}
In this paper, we only consider correlation functions
which are independent of space-time coordinates so that
the space-time volume always factors out.
We use the subscript $density$ to denote
a quantity divided by the volume factor of space-time.
We use this notation for both inner products of string fields
and CFT correlation functions:
\begin{eqnarray}
  \vev{ \Psi_1 | \Psi_2 }
  &=& \int d^{26} x \vev{ \Psi_1 | \Psi_2 }_{density},
\nonumber \\
  \vev{ {\cal O}_1 (z_1) {\cal O}_2 (z_2)
  \cdots {\cal O}_n (z_n) }
  &=& \int d^{26} x \vev{ {\cal O}_1 (z_1) {\cal O}_2 (z_2)
  \cdots {\cal O}_n (z_n) }_{density}.
\end{eqnarray}

The normalization of a state $\ket{\phi}$ in the Fock space
is fixed by the condition
that the $SL(2,R)$-invariant vacuum $\ket{0}$
corresponds to the identity operator. From the normalization
of correlation functions (\ref{three-c})
and the standard mode expansion on a unit circle
\begin{equation}
  c_n = \oint \frac{dz}{2 \pi i} \, z^{n-2} \, c(z), \quad
  b_n = \oint \frac{dz}{2 \pi i} \, z^{n+1} \, b(z),
\end{equation}
the normalization of the inner product is then fixed as follows:
\begin{equation}
  \vev{ 0 | c_{-1} c_{0} c_{1} | 0 }_{density} = 1.
\end{equation}

\section{$\langle \Psi^{(2)} | \Psi^{(2)} \ast \Psi^{(2)} \rangle$
for an arbitrary $t$}
\setcounter{equation}{0}

The explicit expression of
$\langle \Psi^{(2)} | \Psi^{(2)} \ast \Psi^{(2)} \rangle$
for an arbitrary $t$ is given by
\begin{eqnarray}
  && \langle \Psi^{(2)} | \Psi^{(2)} \ast \Psi^{(2)} \rangle_{density}
  = -\frac{81\,{\sqrt{3}}\,\tilde{x}^3}{64}
  -\frac{9\,{\sqrt{3}}\,\left( 22 + 81\,t^2 \right)
      \,\tilde{x}^2\,\tilde{u}}{64}
\nonumber \\ && \quad {}
  -\frac{{\sqrt{3}}\,\left( 2 + 27\,t^2 \right) \,
      \left( 38 + 81\,t^2 \right) \,\tilde{x}\,\tilde{u}^2}{64}
  -\frac{{\sqrt{3}}\,\left( 2 + 3\,t^2 \right) \,
      {\left( 2 + 27\,t^2 \right) }^2\,\tilde{u}^3}{64}
\nonumber \\ && \quad {}
  +\frac{117\,{\sqrt{3}}\,\left( 10 + 27\,t^2 \right)
      \,\tilde{x}^2\,\tilde{v}}{128}
  +\frac{13\,\left( 10 + 27\,t^2 \right) \,
      \left( 22 + 81\,t^2 \right)
      \,\tilde{x}\,\tilde{u}\,\tilde{v}}{64\,{\sqrt{3}}}
\nonumber \\ && \quad {}
  +\frac{13\,\left( 2 + 27\,t^2 \right) \,
      \left( 10 + 27\,t^2 \right) \,
      \left( 38 + 81\,t^2 \right) \,\tilde{u}^2\,\tilde{v}}
      {1152\, {\sqrt{3}}}
\nonumber \\ && \quad {}
  -\frac{13\,\left( 2324 + 7020\,t^2 + 9477\,t^4 \right)
      \,\tilde{x}\,\tilde{v}^2}{256\,{\sqrt{3}}}
\nonumber \\ && \quad {}
  -\frac{13\,\left( 22 + 81\,t^2 \right) \,
      \left( 2324 + 7020\,t^2 + 9477\,t^4 \right)
      \,\tilde{u}\,\tilde{v}^2}{6912\,{\sqrt{3}}}
\nonumber \\ && \quad {}
  +\frac{13\,\left( 167608 + 815724\,t^2 + 1232010\,t^4 + 
      1108809\,t^6 \right) \,\tilde{v}^3}{13824\,{\sqrt{3}}}
\nonumber \\ && \quad {}
  -\frac{9\,{\sqrt{3}}\,\left( 8 + 27\,t^2 \right)
      \,\tilde{x}^2\,\tilde{w}}{32}
  +\frac{3\,{\sqrt{3}}\,\left( 8 - 81\,t^2 \right) \,
      \left( 2 + 3\,t^2 \right)
      \,\tilde{x}\,\tilde{u}\,\tilde{w}}{16}
\nonumber \\ && \quad {}
  +\frac{\left( 38 + 81\,t^2 \right) \,
      \left( 592 - 810\,t^2 - 2187\,t^4 \right)
      \,\tilde{u}^2\,\tilde{w}}{864\,{\sqrt{3}}}
  +\frac{13\,\left( 8 + 27\,t^2 \right) \,
      \left( 10 + 27\,t^2 \right)
      \,\tilde{x}\,\tilde{v}\,\tilde{w}}{32\,{\sqrt{3}}}
\nonumber \\ && \quad {}
  -\frac{13\,\left( 2 + 3\,t^2 \right) \,
      \left( 10 + 27\,t^2 \right) \,
      \left( 8 - 81\,t^2 \right)
      \,\tilde{u}\,\tilde{v}\,\tilde{w}}{96\,{\sqrt{3}}}
\nonumber \\ && \quad {}
  -\frac{13\,\left( 8 + 27\,t^2 \right) \,
      \left( 2324 + 7020\,t^2 + 9477\,t^4 \right)
      \,\tilde{v}^2\,\tilde{w}}{3456\,{\sqrt{3}}}
\nonumber \\ && \quad {}
  -\frac{{\sqrt{3}}\,\left( 4 + 9\,t^2 \right) \,
      \left( 4 + 27\,t^2 \right) \,\tilde{x}\,\tilde{w}^2}{16}
  +\frac{\left( 4 + 27\,t^2 \right) \,
      \left( 376 - 1566\,t^2 - 2187\,t^4 \right)
      \,\tilde{u}\,\tilde{w}^2}{432\,{\sqrt{3}}}
\nonumber \\ && \quad {}
  +\frac{13\,\left( 4 + 9\,t^2 \right) \,
      \left( 4 + 27\,t^2 \right) \,
      \left( 10 + 27\,t^2 \right)
      \,\tilde{v}\,\tilde{w}^2}{288\,{\sqrt{3}}}
  -\frac{{\sqrt{3}}\,t^2\,
      {\left( 4 + 9\,t^2 \right) }^2\,\tilde{w}^3}{8}.
\end{eqnarray}


\renewcommand{\baselinestretch}{0.87}

\begingroup\raggedright\endgroup
\end{document}